\RequirePackage{fix-cm}
\documentclass[twocolumn,epjc3]{svjour3}  
\smartqed  

\RequirePackage{graphicx}
\RequirePackage{amsmath}
\RequirePackage{mathptmx}      
\RequirePackage{flushend}
\RequirePackage[numbers,sort&compress]{natbib}
\RequirePackage[colorlinks,citecolor=blue,urlcolor=blue,linkcolor=blue]{hyperref}
\RequirePackage{bm}
\usepackage{subfigure}
\usepackage{amssymb}

\journalname{Eur. Phys. J. C}
\emergencystretch=\maxdimen

\begin{document}

\title{A femtoscopic Correlation Analysis Tool using~the~Schr\"odinger~equation~(CATS)
}


\author{D.L. Mihaylov\thanksref{e1,addr1}
        \and
        V. Mantovani Sarti\thanksref{e2,addr1} 
        \and
        O.W. Arnold\thanksref{addr1}
        \and
        L. Fabbietti\thanksref{addr1,addr2}
        \and
        B. Hohlweger\thanksref{addr1}
        \and
        A.M. Mathis\thanksref{addr1}
}

\thankstext{e1}{e-mail: dimitar.mihaylov@mytum.de}
\thankstext{e2}{e-mail: valentina.mantovani-sarti@tum.de}


\institute{
Physik  Department T70, E62, Technische Universit\"at M\"unchen,
James  Franck  Str.  1,  85748  Garching,  Germany\label{addr1}
\and~
Excellence Cluster ’Origin and Structure of the Universe’, 
Boltzmannstr. 2, 85748  Garching,  Germany\label{addr2}
}

\date{Received: date / Accepted: date}

\maketitle

\begin{abstract}
We present a new analysis framework called ``Correlation Analysis Tool using the Schr\"odinger equation'' (CATS) which computes the 
two-particle femtoscopy correlation function $C(k)$, with $k$ being the relative momentum for the particle pair. 
Any local interaction potential and emission source function can be used as an input and the wave function is evaluated exactly. 
In this paper we present a study on the sensitivity of $C(k)$ to the interaction
potential for different 
particle pairs: p-p,  p-$\mathrm{\Lambda}$, $\mathrm{K^-}$-p, $\mathrm{K^+}$-p, p-$\mathrm{\Xi}^-$ and $\mathrm{\Lambda}$-$\mathrm{\Lambda}$. 
For the p-p Argonne $v_{18}$ and Reid Soft-Core potentials have been tested. For the other pair systems we present 
results based on strong potentials obtained from effective Lagrangians such as $\chi$EFT for p-$\mathrm{\Lambda}$,
J\"ulich models for $\mathrm{K(\bar{K})}$-N and Nijmegen models for $\mathrm{\Lambda}$-$\mathrm{\Lambda}$. 
For the p-$\mathrm{\Xi}^-$ pairs we employ the latest lattice results from the HAL QCD collaboration.
\\
Our detailed study of different interacting particle pairs as a function of the source size 
and different potentials shows that femtoscopic measurements 
can be exploited in order to constrain the final state interactions among hadrons.
\\
In particular, small collision systems of the order of 1~fm, as produced in pp collisions at the LHC, seem to provide a 
suitable environment for quantitative studies of this kind.
\end{abstract}

\section{Introduction}
\label{intro}
Femtoscopy has been mainly employed so far to study the properties of the particle emitting source in heavy-ion collisions 
by analyzing particle pairs with low relative momentum undergoing a known interaction ~\cite{Pratt:1984su, Lisa:2005dd}. 
In heavy-ion collisions pion femtoscopy has been the most common tool to get insight 
into the time-space evolution of the produced medium~\cite{Agakishiev:2011zz, Kotte:2004yv,
Aggarwal:2007aa, Adams:2004yc, Aamodt:2011mr, Adams:2005ws, Anticic:2011ja, Chung:2002vk, Agakishiev:2010qe, Adamczyk:2014vca, Adam:2015vja}.
\\
Unlike heavy-ion collisions, nucleon-nucleon (NN) collisions are not affected by the formation of a medium such as the Quark-Gluon-Plasma. 
This means that the time evolution of the source can be neglected and the hadron interaction is not influenced by collective processes.
\\
More generally, femtoscopy allows to describe any hadron-hadron correlation involving both (anti)mesons and (anti)baryons. 
Recent femtoscopy studies in pp at $\sqrt{s}=7$~TeV and p-Pb at $\sqrt{s_\mathrm{NN}}=5.02$ TeV showed that in these systems a smaller 
source size radius is extracted compared to heavy-ion collisions~\cite{PhysRevD.84.112004,Abelev:2012sq, PhysRevC.91.034906}. 
When mesons are considered the correlation function is affected by a mini-jet background over the whole $k$ range~\cite{PhysRevD.84.112004,Abelev:2012sq}. 
This effect is also visible in baryon-antibaryon correlations ($\mathrm{B}$-$\mathrm{\bar{B}}$), where partons from the colliding 
protons hadronize in particle cones containing a $\mathrm{B}$-$\mathrm{\bar{B}}$ pair. The production of such pairs is enhanced due to 
the intrinsic baryon number conservation and therefore shows a strong kinematic correlation which overlaps to the true femtoscopic signal in the final correlation function. 
Since these mini-jets are not present in the correlation function of $\mathrm{B}$-$\mathrm{B}$/$\mathrm{\bar{B}}$-$\mathrm{\bar{B}}$~\cite{Adam:2016iwf}, those systems 
are well suited to study the final state interaction.
\\
The femtoscopic formalism includes the computation of the correlation function for a given source function and interaction potential~\cite{Lisa:2005dd}
\begin{align}\label{corr}
C(k)=\int S(\vec{r}) \mid \psi_{k}(\vec{r})\mid ^2\mathrm{d}^3 r,
\end{align}
where $k=(\mid \mathbf{p}_1-\mathbf{p}_2\mid)/2$ is the reduced relative momentum in the center of mass of the pair ($\bf{p}_1+\bf{p}_2=0$), $\vec{r}$ is the 
relative distance between the two particles, $S(\vec{r})$ is the source function and $\psi_{k}(\vec{r})$ is the two-particle wave function.
\\
The above formula is based on the assumptions 
that space and momentum emission coordinates are uncorrelated and the emission process is time independent~\citep{Lisa:2005dd}. 
The dynamics of the processes is explicitly embedded on one hand in the emitting source $S(\vec{r})$, which depends on the colliding system, and on the other 
hand in the interaction potential between the two particles expressed by the relative wave function $\psi_{k}(\vec{r})$.
\\~\\
The source could be either parametrized using a specific distribution function, e.g. a Gaussian or a Cauchy function, or extracted directly from transport models 
such as EPOS~\cite{PhysRevC.92.034906}.
The pair wave function can be obtained by solving the Schr\"odinger equation for a specific interaction potential and a popular tool to do 
that is the Correlation Afterburner (CRAB)~\cite{crab}. 
This afterburner only delivers the asymptotic solution at large relative distances which eventually leads to a wrong correlation function 
at small distances. For heavy-ion collisions 
the source radius is typically between 2 and 6~fm~\cite{Agakishiev:2011zz, Kotte:2004yv, Aggarwal:2007aa, Adams:2004yc, Aamodt:2011mr}. 
However in pp collisions at LHC energies the extracted source can be even smaller than 1.5~fm~\cite{PhysRevD.84.112004,Abelev:2012sq, ourpaper}, and 
the EPOS transport model predicts a non-Gaussian source which peaks at distances of around 1~fm (see Fig.~\ref{fig:EposSource}).
\\
This calls for a tool capable of providing an exact solution of the Schr\"odinger equation valid also at short distances.
\\~\\
Another common approach based on solving the Schr\"odinger equation and used to study p-p correlations is the 
Koonin model~\cite{Koonin:1977fh}. Here a Gaussian source distribution is assumed and the relative wave function is obtained by 
employing a Coulomb potential and a Reid-Soft Core potential for the strong interaction~\citep{Reid:1968sq}.
\\
Analytical solutions to compute $C(k)$ do exist, but they are limited by different approximations.
One of the most popular analytical models was developed by Lednick\'{y} and Lyuboshits~\citep{Lednicky:1981su}. 
The emission source is assumed to have a Gaussian shape and the wave function is modeled within the effective range approximation, 
using the scattering length $(a_0)$ and the effective range of the potential $(r_{eff})$. 
Such a parametrization is only sensitive to the asymptotic region of the interaction. 
Due to its long range nature, the Coulomb potential is normally not considered in the Lednick\'{y} model and can at most 
be treated in an approximate way ~\cite{Adam:2015vja}. 
\\~\\
All the above-mentioned issues motivated the development of a new femtoscopy tool called 
``Correlation Analysis Tool using the Schr\"odinger equation'' (CATS).
\\ 
CATS is capable of evaluating numerically the full wave function without using any approximations and computing $C(k)$ for different source functions and interaction potentials. 
CATS is faster than most of the conventional numerical tools and also more flexible since it can be used together with an external fitter.
\\~\\
This paper is organized as follows. In Section~\ref{overview} we introduce the 
formalism of the Schr\"odinger equation as embedded in the algorithm, while in Section~\ref{interaction} we introduce the interaction potentials 
used for the later investigations of the p-p, p-$\mathrm{\Lambda}$, $\mathrm{K}$($\mathrm{\bar{K}}$)-p, p-$\mathrm{\Xi}^-$ and $\mathrm{\Lambda}$-$\mathrm{\Lambda}$ systems.
\\
In Section~\ref{results} we show our results for the correlation functions of the above-mentioned pairs using different interaction potentials 
and source sizes. A comparison with the emitting source obtained from the EPOS transport model is presented for p-p and p-$\mathrm{\Lambda}$ pairs. 
In addition we show results for p-$\mathrm{\Xi}^-$ and $\mathrm{\Lambda}$-$\mathrm{\Lambda}$ correlation 
functions including experimental effects such as momentum resolution and feed-down decays. 
Moreover we present a comparison between the extracted source radius for p-p pairs obtained from the HADES collaboration in p-Nb reactions 
at $\sqrt{s_\textrm{NN}}=3.18$ GeV and a CATS fit of the same data. 
\\
Finally in Section~\ref{conclusion} we present conclusions and future outlooks.
In addition in \ref{appA} we provide technical information related to the installation of CATS and the numerical methods used in the code.

\section{CATS}
\subsection{Overview}\label{overview}
\begin{figure}[h]
\centering{
\includegraphics[width=\columnwidth]{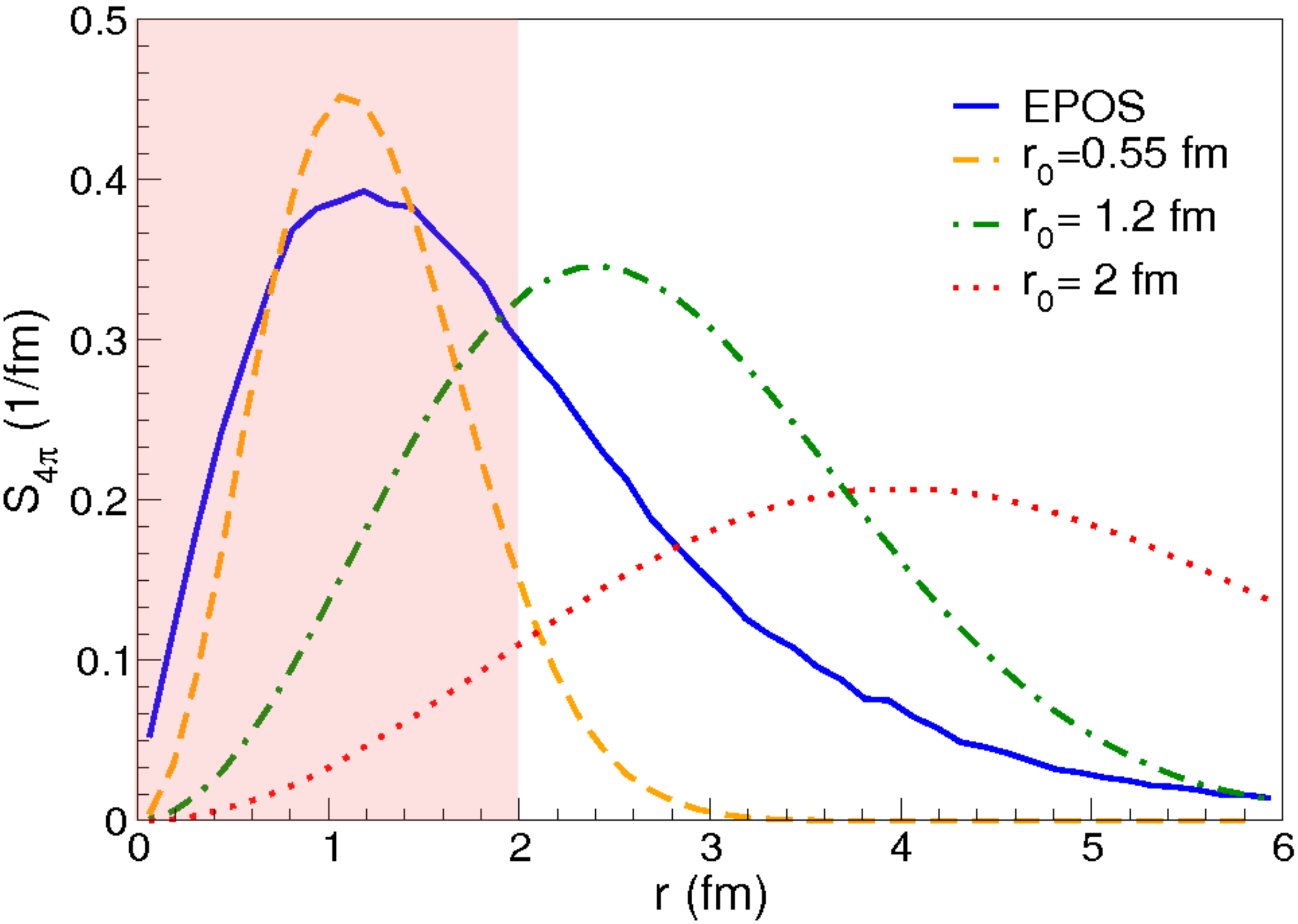}
}
\caption[The EPOS source]
{The probability density function of the relative distance $r$ for p-p pairs produced 
in pp collisions at 7~TeV. 
The blue solid line is the prediction from the EPOS transport model, the dashed orange line, the green dot-dashed line and the dotted red line are 
Gaussian sources with different radii. For better visibility the source with $r_0=0.55$~fm is scaled down by 0.6. 
The red shaded area shows the typical range of the strong interaction where an approximate solution to the Schr\"odinger equation may fail.
}
\label{fig:EposSource}
\end{figure}

CATS is  
designed to compute the two-particle correlation function for any emission distribution and interaction potential. 
This is achieved by numerically solving the Schr\"odinger Equation (SE) and by evaluating the 
convolution of the resulting wave function with a source distribution (see Eq.~(\ref{corr})).
\\
The two-particle stationary SE reads
\begin{align}\label{eq:SE}
-\dfrac{\hbar ^2}{2\mu}\nabla ^2 \psi +V\psi=E\psi,
\end{align}
where $\mu=\frac{m_1m_2}{m_1+m_2}$ is the reduced mass of the system.
\\
By assuming a central interaction potential $V$, the total wave function is separated into a radial term and an angular term given by the spherical harmonics
\begin{align}
\psi_k(r,\theta,\phi)=R_k(r)Y_l ^m (\theta,\phi). 
\end{align}
In particular the radial equation to be solved is
\begin{align}
\label{eq:SE}
-\dfrac{\hbar^2}{2\mu}\dfrac{\mathrm{d}^2 u}{\mathrm{d} r^2}+\left[V(r)+\dfrac{\hbar^2}{2\mu}\dfrac{l(l+1)}{r^2}\right] u = E u,
\end{align}
where $u(r)=r R(r)$.
\\
The overall interaction potential $V(r)$ is the sum of a short range strong contribution and a long range Coulomb potential. 
Since we are looking for scattering states as a function of the relative momentum $k$ and not for bound states, the energy $E$ of the system needs to be fixed by using the 
relation $E=\frac{\hbar^2 k^2}{2\mu}$. The scattered state will then approach the asymptotic solution outside the range 
of the strong potential leading to a free or a pure Coulomb wave function depending on the charge of the particles involved.
\\ In order to properly match the 
exact solution to the asymptotic form we have to evaluate the phase shifts of the corresponding wave function.
For this purpose the SE is solved by expanding the total wave function in partial waves 
\begin{align}
\psi_k(r,\theta,\phi)&=\sum_{l=0} ^{l_{max}}R_{k,l}(r) Y_l ^m (\theta,\phi)\nonumber \\
&=\sum_{l=0} ^{l_{max}} i^l (2l+1)\dfrac{u_{k,l}(r)}{r} P_l (\cos \theta).
\end{align}
where $P_l (\cos \theta)$ are the Legendre polynomials.
\\
The sum over the partial waves runs 
until the convergence of the solution is reached, and from there we obtain the total wave function $\psi_{k}(\vec{r})$ to be used in Eq.~(\ref{corr}).
\\
Taking into account the Fermi statistics of the particle pairs and fixing the isospin ($I$) configuration, only specific states $^{2S+1}L_J$ are allowed depending 
on the particle species, where we use the standard spectroscopic notation with $S$ denoting the total spin, $L$ the orbital angular momentum 
and $J$ the total angular momentum of the pair.
Moreover, since NN, nucleon-hyperon (NY), kaon-nucleon (KN) and hyperon-hyperon (YY) potentials 
might involve spin-orbit coupling $\bar{L}\cdot \bar{S}$, the total angular momentum $\bar{J}=\bar{L}+\bar{S}$ 
has to be taken as a good quantum number to characterize the eigenstates
and has to be accounted for in the total degeneracy of the states.
\\
The total correlation function that CATS provides as an output is then given by
\begin{align}
C(k)=\sum _{\textrm{states} (I,S,L,J)} w_{(I,S,L,J)} C_{(I,S,L,J)},
\end{align} 
where each $C_{(I,S,L,J)}$ is evaluated by means of Eq.~(\ref{corr}) and weighted by $w_{(I,S,L,J)}$.
\\
In experimental conditions with unpolarized particles we should take into account the degeneracy in $S$, 
as well as in $I$. Moreover for $L>0$ states 
also the degeneracy in $J$ has to be considered. How to compute the corresponding weights $w_{(I,S,L,J)}$ is explained in \ref{appA}.
\\~\\
In CATS the source function is characterized in two ways. One possibility is to define an analytical function which 
models the emission source. It is typically based on a Gaussian 
distribution of the $x$,$y$ and $z$ single particle coordinates. 
The emission source $S(r,\theta,\phi)$ is defined as the probability density function (pdf) 
to emit a particle pair at a certain relative distance $r$ and relative polar and azimuthal angles $\theta$ and $\phi$.
For an uncorrelated particle emission the Gaussian source reads
\begin{equation}\label{eq:Gauss}
 S(r) = \frac{1}{(4\pi r_0 ^2)^{3/2}}\exp{\left(-\dfrac{r^2}{4r_0 ^2}\right)},
\end{equation}
where $r_0$ is the size of the source. Since no angular dependence is involved, to obtain 
the probability of emitting two particles at a distance $r$ a trivial integration over the solid angle is necessary
\begin{equation}\label{eq:Gauss4Pi}
 S_{4\pi}(r) = 4\pi r^2S(r)= \frac{4\pi r^2}{(4\pi r_0 ^2)^{3/2}}\exp{\left(-\dfrac{r^2}{4r_0 ^2}\right)}.\\
\end{equation}
Another possibility is to sample the relative distance between the particle pairs directly
from the output of an event generator such as EPOS.
\\
In Fig.~\ref{fig:EposSource} we compare the shapes of differently 
sized Gaussian sources to the EPOS source for p-p pairs obtained in a simulation of pp collisions at 7~TeV. 
Notably the EPOS source predicts that most of the particle pairs will be emitted at distances below 2~fm, which is the typical range 
of the strong interaction. Hence the asymptotic solution of the wave function will no longer be valid in that region, highlighting 
the necessity of an exact treatment of the problem. 
Moreover we see that EPOS predicts a non-Gaussian source.

\subsection{NN and NY interaction potentials}\label{interaction}
The NN interaction has been widely constrained  thanks to a large amount of scattering data along with data on bound states such as the 
deuteron~\cite{Stoks:1993tb}.
In the common Yukawa picture the NN potential is described by means of a long range 
One-Pion-Exchange term (OPE), a $\pi\pi$ exchange term for the intermediate attraction and a vector meson ($\omega$) exchange accounting for the repulsive core.
\\
This repulsive contribution can be phenomenologically modeled with a Woods-Saxon function.
\begin{figure}[h]
\centering{
\includegraphics[width=\columnwidth]{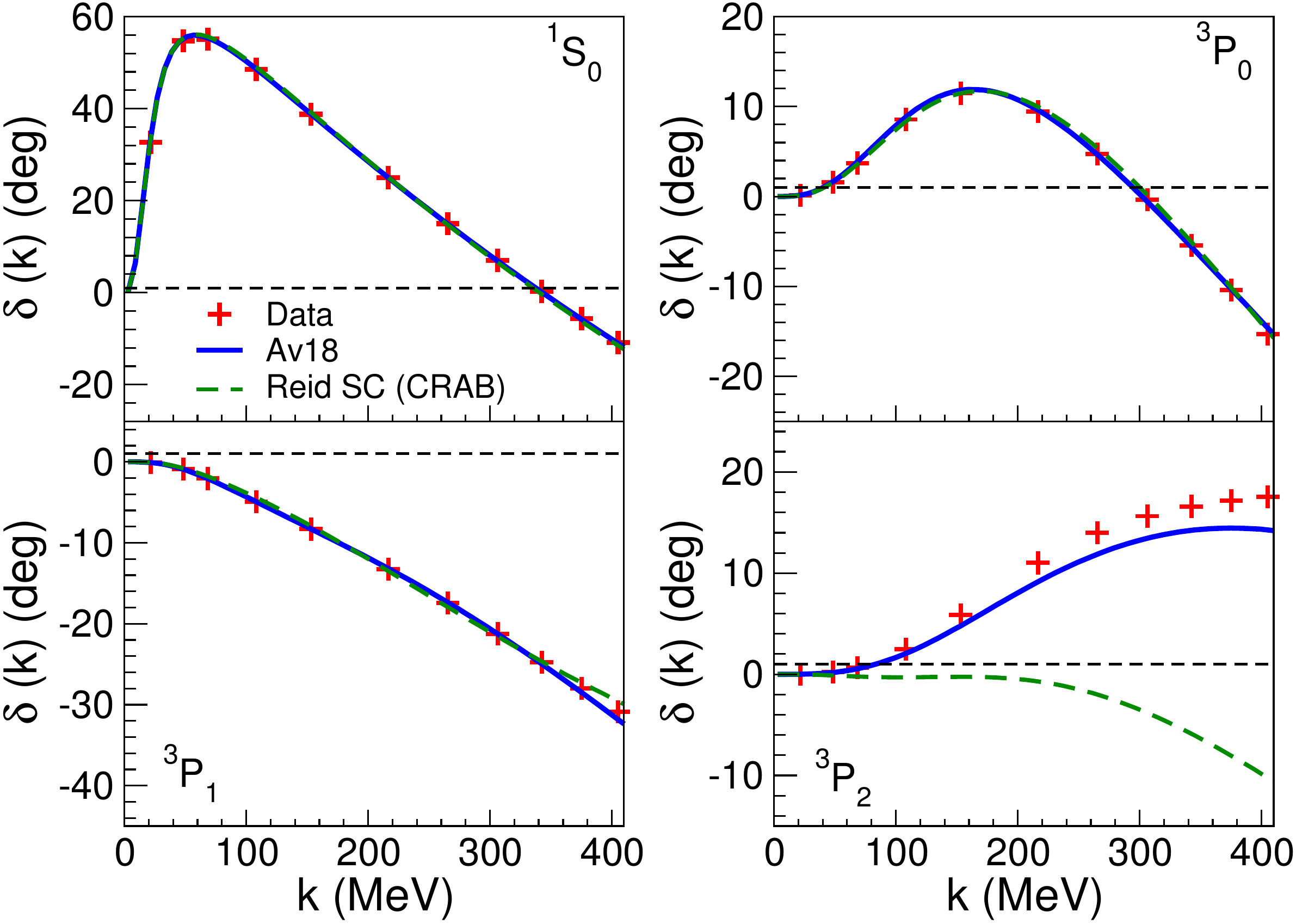}
}
\caption{Phase shifts as a function of the relative momentum $k$ for the p-p system using  the Argonne $v_{18}$ potential 
(blue solid line) and the Reid SC potential as used in CRAB calculations (green dashed line). 
The experimental data are taken from ~\cite{Stoks:1993tb}.}
\label{fig:ppphaseshift}
\end{figure}
All these features are included in the most common NN potentials such as Bonn~\cite{Machleidt:1987hj,Machleidt:1989tm}, 
Urbana~\cite{Lagaris:1981mm}, Paris~\cite{Lacombe:1980dr}, Nijmegen~\cite{Nagels:1977ze}, Reid~\cite{Reid:1968sq} and Argonne~\cite{Wiringa:1994wb} potentials.
In addition, the results obtained within a chiral effective field theory approach ($\chi$EFT)~\cite{ Entem:2015xwa, Epelbaum:2014sza, Entem:2003ft, Epelbaum:1998ka, Kaiser:1997mw} 
and within the latest version of the Nijmegen model ESC08~\cite{Nagels:2014qqa} have to be mentioned, since they provide  similar results for the scattering parameters and 
both models show a nice description of the available data.
\\~\\
For the study of the p-p correlation function in this paper we employ the Argonne $v_{18}$~\cite{Wiringa:1994wb} and the Reid Soft-Core (RSC)~\cite{Reid:1968sq} 
potentials. The latter is the default choice in femtoscopic analyses~\cite{Koonin:1977fh,crab}. In the Koonin model only the RSC singlet state $^1 S_0$ is considered, 
while in the CRAB calculations also the triplet $^3 P_2$ state is included but without spin-orbit or tensor terms.
Nevertheless, these terms are crucial in order to reproduce the $^3 P_2$ phase shift, 
as the Argonne $v_{18}$ potential shows a better agreement to pp scattering data in the momentum range we are interested in (see Fig.~\ref{fig:ppphaseshift}).
\begin{figure}[h]
\centering{
\includegraphics[width=\columnwidth]{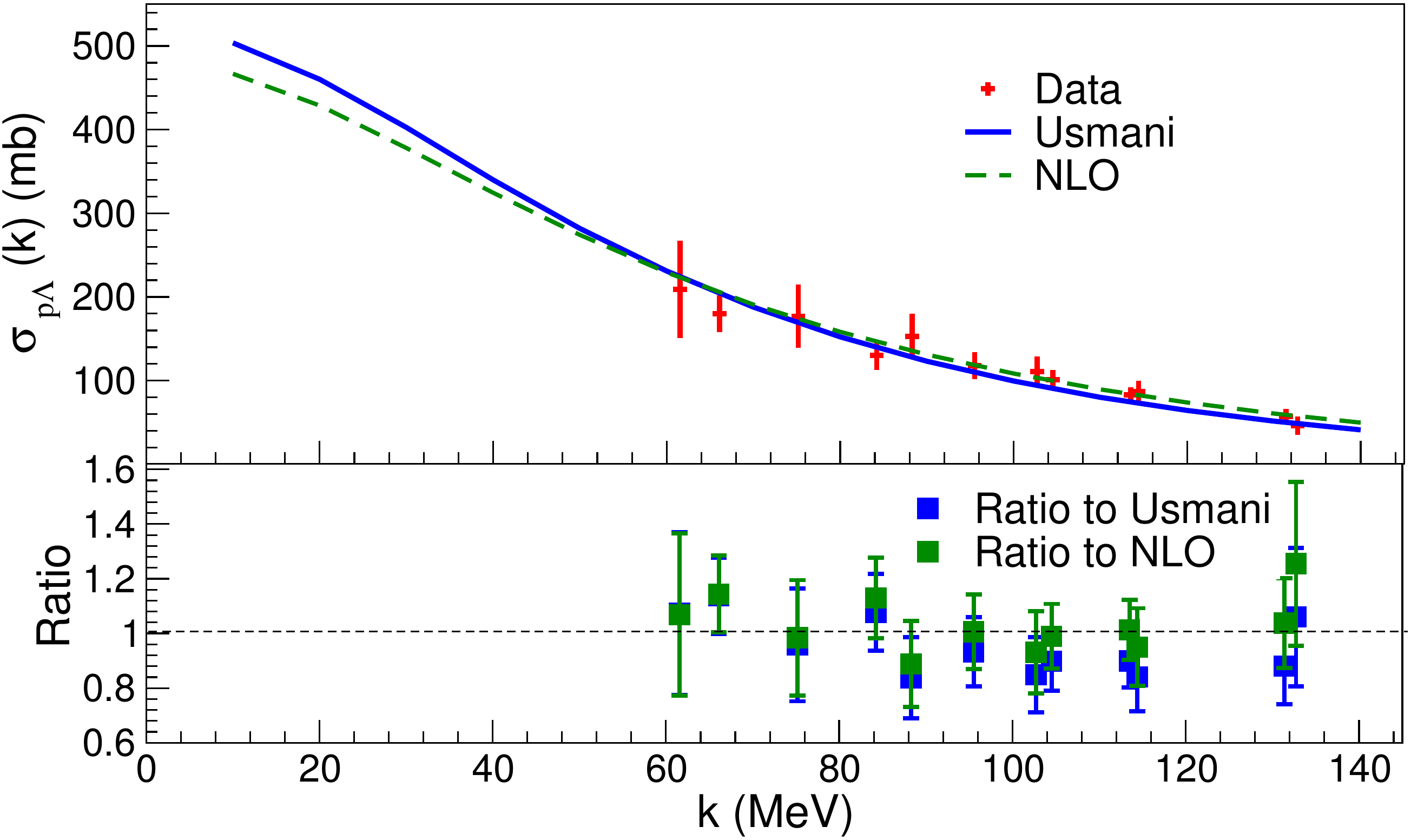}
}
\caption{Cross section for p-$\mathrm{\Lambda}$  as a function of the relative momentum $k$ evaluated in CATS using the Usmani (blue solid line) and NLO $\chi$EFT 
(green dashed line) potentials. Experimental data are taken from~\cite{PhysRev.173.1452,PhysRev.175.1735}.}
\label{fig:aa}
\end{figure}
\\
The range of relative momenta $k$ relevant for femtoscopic studies stops at $k_{max}\approx 200$ MeV/c and the dominant contribution in 
the phase shifts in this regime is mainly coming from s- and p-wave states~\cite{Stoks:1993tb}.
For this reason at the moment we are neglecting the coupled channel to the $^3 F_2$ wave.
\\~\\
Unfortunately in the nucleon-hyperon sector only $36$ scattering data points have been measured~\cite{pap54,Eisele:1971mk,Hepp:1968zza,pap57,pap58,Bart:1999uh} 
and the available phenomenological potentials are fitted in order to reproduce these data along with the accessible hypernuclei binding energies.
In particular, for the p-$\mathrm{\Lambda}$ interaction we adopt the potential introduced
by Bodmer, Usmani, and Carlson~\cite{PhysRevC.29.684}, which reproduces the available cross section data as shown in Fig.~\ref{fig:aa}. 
This potential is similar to a NN Argonne interaction since it includes a repulsive Woods-Saxon core and a two-pion intermediate exchange term.
At the moment we just consider s-waves, which account for both the singlet $^1S_0$ and the triplet $^3S_1$ state.
\\
To test the sensitivity of the correlation function to the p-$\mathrm{\Lambda}$ interaction, we have also used the results obtained by next-to-leading order (NLO) 
$\chi$EFT calculations~\cite{Haidenbauer:2013oca}. In this work the authors showed that the agreement with the available data in the $S=-1$ sector is 
considerably improved with respect to previous leading order results. A local potential  for this model is not yet available, nevertheless we were provided with the 
radial wave function and used CATS to compute the resulting correlation functions for different sources. The scattering parameters 
of the two potentials used to model the p-$\mathrm{\Lambda}$ interaction are listed in Table \ref{tabNloUsmani}.
\begin{center}
\begin{table}[h]
\caption{The p-$\mathrm{\Lambda}$ scattering parameters of the Usmani and the chiral potentials. 
We denote with $(s)$ and $(t)$ the singlet and triplet state respectively. The scattering parameters 
for the Usmani potential are evaluated here \cite{PhysRevC.29.684}. The parameters for the 
the chiral potential are taken from \cite{Haidenbauer:2013oca} for a cutoff scale of 600~MeV.
}
\label{tabNloUsmani}
\centering
\begin{tabular}{ c c c }
\hline 
\textbf{~Parameter~} & \textbf{~Usmani~} & $\mathbf{\chi}$\textbf{EFT NLO}\\ 
\hline 
$a^{(s)}_0$~(fm) & $-2.88$ & $-2.91$\\ 
\hline 
$r^{(s)}_{eff}$~(fm) & $2.92$ & $2.78$\\
\hline
$a^{(t)}_0$~(fm) & $-1.66$ & $-1.54$\\ 
\hline 
$r^{(t)}_{eff}$~(fm) & $3.78$ & $2.72$\\
\hline
\end{tabular} 
\end{table}
\end{center}
Concerning the $S=-2$ sector the interaction of $\mathrm{\Xi}$ hyperons with nuclei is not well constrained experimentally or phenomenologically~\cite{Gal:2016boi}.
Recently a $\mathrm{\Xi}$-hypernucleus candidate has been detected~\cite{Nakazawa:2015joa} and ongoing measurements suggest 
that the N-$\mathrm{\Xi}$ interaction is weakly attractive~\cite{Nagae:2017slp}.
\\
Recent lattice QCD calculations from the HAL QCD collaboration showed preliminary results on the p-$\mathrm{\Xi}^-$ correlation 
function based on local potentials in the s-wave singlet and triplet states, both in $I=0$ and $I=1$ channels~\cite{Hatsuda:2017uxk, Sasaki:2017ysy}.
The interaction potentials have been obtained in $(2+1)$-flavor lattice QCD simulations close to the 
physical masses of pions and kaons. In particular in the isoscalar $^1 S_0$ channel it has been shown 
that the p-$\mathrm{\Xi}^-$ interaction is deeply attractive at intermediate distances and relatively weakly 
repulsive at shorter distances. The isoscalar spin triplet state $^3 S_1$ has a shallow 
attractive well and a very repulsive core.
\\
The $I=1$ channel does not present any attraction in the spin singlet state while a mildly attractive region enclosing a 
repulsive inner core has been found in the spin triplet state. These contributions might not be relevant for large source sizes~\cite{Hatsuda:2017uxk} 
but could impact the overall correlation function for smaller emitting sources as obtained in elementary collisions.

\subsection{KN and $\mathrm{\bar{K}N}$ interaction}\label{interactionk}
As for the hyperon-nucleon case, the (anti)kaon-nucleon system is a powerful tool to study the short range strong interaction with the exchange of 
strangeness as a new degree of freedom.
\\
In particular kaons are suited to study the inner part of nuclei since at low energies they 
interact rather weakly with the nucleons~\cite{Gal:2016boi}. On the contrary antikaons scattering on nuclei is characterized by 
absorption processes that can lead to the production of hyperons such as $\mathrm{\Lambda}$ and $\mathrm{\Sigma}$~\cite{katz,Gal:2016boi}.
\\
The understanding of the short range kaon-nucleon interaction, already in vacuum, is fundamental to unveil more complicated 
systems such as quasi-bound states near threshold, kaonic atoms and kaonic 
clusters~\cite{PhysRevLett.2.425,Chao:1973sa,Engler:1965zz,Agakishiev:2012xk,Moriya:2009mx,Bazzi:2011zj,Agnello:2005qj}.
\\
From a theoretical point of view, based on the successful description of the NN interaction, several meson-baryon exchange 
models, such as the J\"{u}lich model~\cite{MuellerGroeling:1990cw,Buettgen:1990yw}, have been applied so far in the description of the available KN data. 
In this approach the short range repulsion in $\mathrm{K^+}$-N is not well reproduced by the simple presence of the $\omega$ meson. 
In order to reproduce the s-wave scattering parameters an additional repulsive contribution is needed. The inclusion of spin-dependent 
terms, such as one gluon exchange (OGE) or exchange of the scalar-isovector  meson $a(980)$, improves the agreement in the s channel~\cite{Hadjimichef:2002xe}.
\\
It is also worth mentioning that different theoretical approaches based on quark models~\cite{Bender:1983cw,SilvestreBrac:1997jw,Lemaire:2002wh} and $\chi$EFT 
interactions~\cite{Ikeda:2012au} have also been used to investigate the kaon-nucleon interaction.
\\
In this work we adopted the above mentioned J\"{u}lich model, which besides single particle exchange terms also contains box diagrams allowing for intermediate 
states such as $\mathrm{N},\mathrm{\Delta},\mathrm{K(\bar{K})},\mathrm{K^\ast(\bar{K^\ast})}$. A local potential was not available for this model, nevertheless 
we were provided with the exact solution of the radial wave function, which was included in CATS for the computation of the correlation function.
\\ 
Besides the wave function corresponding to the strong K($\mathrm{\bar{K}}$)-N interaction potential, we also take into account the Coulomb interaction
in an approximate way by multiplying the (strong) wave function with the Gamow factor $A_c(\eta)$~\cite{Adam:2015vja}
\begin{align}
&\psi_\mathrm{KN,\bar{K}N}(r,\theta,\phi)=A_c(\eta) \psi^{s}(r,\theta,\phi)\nonumber\\
&A_c(\eta)=2\pi\eta(e^{2\pi\eta}-1)^{-1},
\end{align}
where $\eta=\left(\frac{1}{2}a_s k_\mathrm{KN,\bar{K}N}\right)^{-1}$ and $a_s=(m_{red}z_1 z_2 e^2)^{-1}$, with $z_{1,2}$ being the charge numbers of the two particles. 
A more elaborate approximation based on the renormalization of the asymptotic behaviour of the strong wave function has been developed 
in the past~\cite{HOLZENKAMP1989485} and will be used in future analyses.

\subsection{YY interaction}\label{interactionyy}
In the YY sector the available experimental data on the binding energy of  $\mathrm{\Lambda\Lambda}$ hypernuclei, 
despite recent data on $\mathrm{ _{\Lambda\Lambda} ^6 He}$~\cite{Takahashi:2001nm, Ahn:2013poa}, do not allow to set 
tight constraints on the nature of the underlying interaction~\cite{Gal:2016boi}.
\\
Recently the STAR collaboration employed the femtoscopy technique to study $\mathrm{\Lambda}$-$\mathrm{\Lambda}$ correlations in Au-Au collisions at 
$\sqrt{s_\textrm{NN}}$ = 200 GeV~\cite{Adamczyk:2014vca}. The reported shallow repulsive interaction in this work 
is still under debate, since in an alternative analysis, that considered the contribution of the residuals in a more sophisticated way, 
a shallow attractive interaction was confirmed~\cite{Morita:2014kza}. 
One of the goals of this work is to show that a detailed study of the $\mathrm{\Lambda}$-$\mathrm{\Lambda}$ correlation function might be a sensitive tool to extract 
information on the interaction of the two particles.
\\
For this purpose, we select four different potentials: 
three meson-exchange Nijmegen models with a short range hard-core (ND56~\cite{Nagels:1976xq}, NF50 and NF42~\cite{Nagels:1978sc}) and one quark-model 
potential which includes meson exchange effects as well (fss2~\cite{Fujiwara:2006yh}).
The models ND56, NF50 and fss2 deliver the best agreement with the $\mathrm{\Lambda}$-$\mathrm{\Lambda}$ correlation data measured by STAR.
The NF42 has been chosen to show the sensitivity of the correlation function to the presence of a possible bound state, which is allowed by this potential.
\begin{figure}[h]
\centering{
\includegraphics[width=\columnwidth]{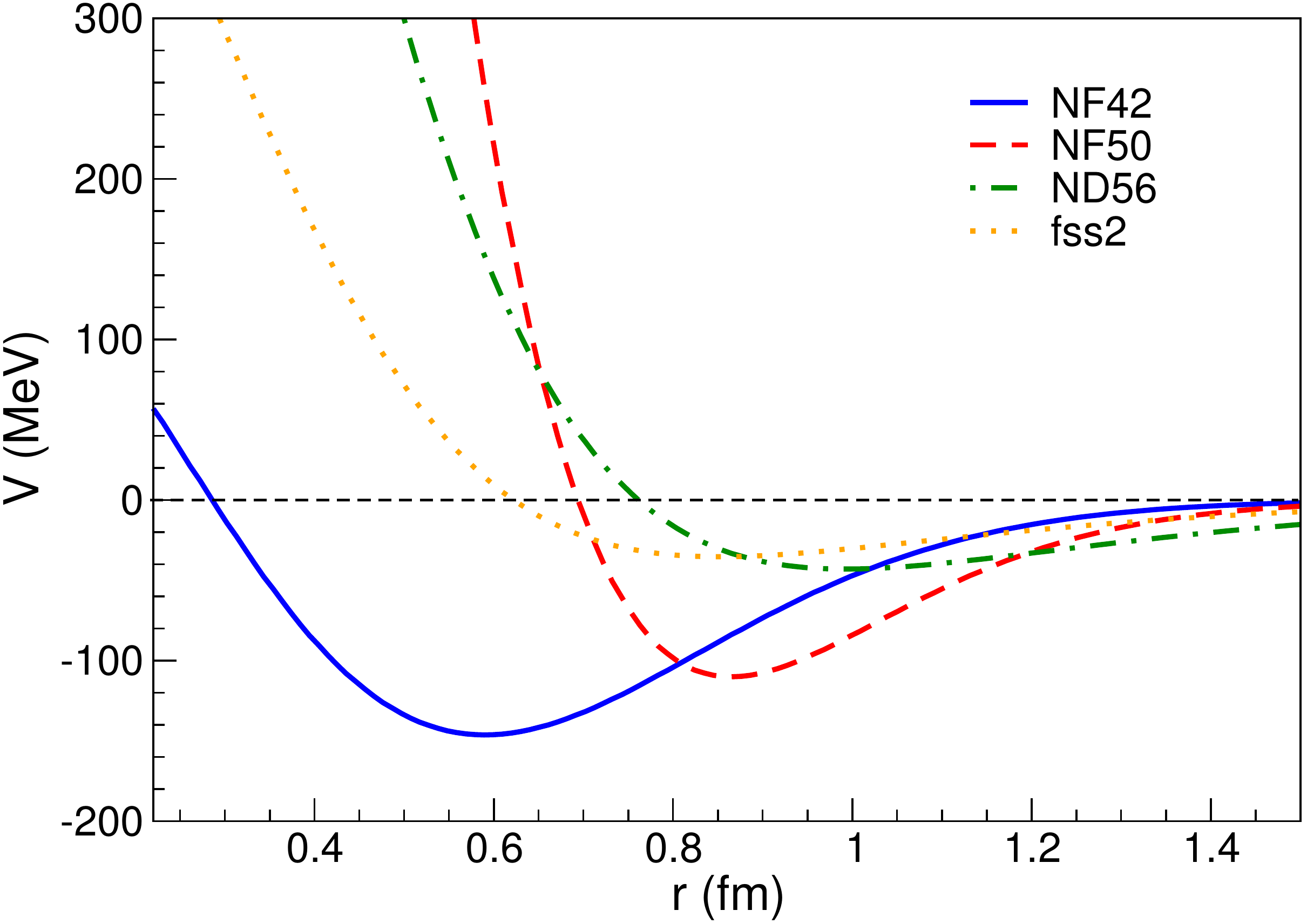}
}
\caption{$\mathrm{\Lambda}$-$\mathrm{\Lambda}$ interaction potentials as a function of the relative distance $r$. 
The parametrization in Eq.~(\ref{vll}) is taken from~\cite{Morita:2014kza}.}
\label{fig:SourceAndPotentialsL}
\end{figure}
\\
The above mentioned potentials have been translated in a local form by using a two-range Gaussian parametrization, following the approach used in~\cite{Morita:2014kza}
\begin{align}\label{vll}
V_{\mathrm{\Lambda\Lambda}}(r)=v_1 e^{-r^2/\mu_1 ^2}+v_2 e^{-r^2/\mu_2 ^2}.
\end{align}
The parameters are fixed to values that reproduce the scattering parameters obtained within each model. They are summarized in Table~\ref{tabyy}. 
In Fig.~\ref{fig:SourceAndPotentialsL} the potentials are plotted as a function of the relative distance $r$.
\\
It is worthwhile mentioning that only the NF42 model provides a positive scattering length $a_0\approx 3.7$~fm due to the 
presence of a bound state, while all the remaining potentials lead to negative (attractive) values of $a_0$ and effective ranges between $4$ and $5$~fm. 

\begin{table*}[t]
\caption{The $\mathrm{\Lambda\Lambda}$ potential parameters used in the two-range Gaussian parametrization in eq.~(\ref{vll}) 
and the corresponding scattering parameters $a_0$ (fm) and $r_{eff}$ (fm). The values are taken from ~\citep{Morita:2014kza}.}\label{tabyy}
\begin{tabular*}{\textwidth}{@{\extracolsep{\fill}}ccccccc@{}}
\hline
\textbf{Model} & $\bm{v_1}$ \textbf{(MeV)} & $\bm{\mu_1}$ \textbf{(fm)} & $\bm{v_2}$ \textbf{(MeV)} & $\bm{\mu_2}$ \textbf{(fm)} & $\bm{a_0}$ & $\bm{r_\textit{\textbf{eff}}}$  \\ 
\hline 
ND56 & $-144.26$ & $1.0$ & $1413.75$ & $0.45$ & $-1.179$ & $4.656$ \\ 
\hline 
NF42 & $-878.97$ & $0.6$ & $1048.58$ & $0.45$ & $3.659$ & $0.975$\\
\hline
NF50 & $-2007.35$ & $0.6$ & $5678.97$ & $0.45$ & $-0.772$ & $4.271$ \\ 
\hline 
fss2 & $-103.9$ & $0.92$ & $658.2$ & $0.41$ & $-0.81$ & $3.99$ \\ 
\hline
\end{tabular*}
\end{table*}

\section{Results}\label{results}

In Fig.~\ref{fig:crab1} a comparison between the correlation functions obtained by 
CRAB and CATS for p-p pairs is shown. In particular two different transport models are 
used to model the source: UrQMD \cite{Bass:1998ca} for simulating p-Nb reactions at $\sqrt{s_{NN}}=3.18$~GeV and 
EPOS~\cite{PhysRevC.92.034906} for simulating pp reactions at $\sqrt{s}=7$~TeV. 
The p-Nb system is expected to have a Gaussian source of around $2$~fm~\cite{Adamczewski-Musch:2016jlh}, 
while the pp system produces a much smaller source, comparable to a Gaussian source of below 1~fm (see Fig.~\ref{fig:EposSource}). 
As already pointed out the CRAB tool provides only an approximate solution at small distances. 
Fig.~\ref{fig:crab1} shows that in the case of the larger source (the p-Nb system) the two models are in agreement within few percent. 
However for pp collisions at 7~TeV the source is much smaller and consequently a much larger discrepancy, of up to 20\%, is observed between CRAB and the 
exact CATS solution.
\\~\\
In Fig.~\ref{fig:res2} we show the correlation function $C(k)$ obtained by CATS for p-p pairs (upper panel) and p-$\mathrm{\Lambda}$ pairs (lower panel) 
with different interaction potentials and Gaussian source sizes.
\\
In particular for p-p pairs we perform the calculations for the RSC potential as used in the Koonin and CRAB models and compare them to the A$v_{18}$ potential. 
The profile of the resulting correlation functions reproduce the main features coming from the interplay between the attractive part of the NN potential 
and the short-range repulsive core along with contributions from Coulomb and Fermi-Dirac statistics. 
The repulsive interaction is visible between $60<k<200$ MeV since $C(k)$ is below unity. This effect is enhanced for smaller sources.
\\
The strength of the p-p correlation signal increases by  almost a factor $2.5$ going from $r_0=2$~fm to $r_0=0.85$~fm.
The differences among the potentials are negligible for most of the source sizes. Nevertheless it is evident that for sources below $1~$fm 
deviations up to 10\% are present at $k>150$~MeV.
\\ 
The p-$\mathrm{\Lambda}$ pair is neither affected by Coulomb interaction nor by quantum statistics. 
As a result the shape of the corresponding correlation function $C(k)$ is only affected by the strong interaction potential. 
Here results are shown for the Usmani potential and the $\chi$EFT NLO potential (see Sec.~\ref{interaction}). 
Both of these potentials possess a phenomenological repulsive core which is dominated by an intermediate attraction as $k$ decreases. 
This results in a rise of $C(k)$ at lower momenta. The two potentials produce very similar correlation functions and the differences are mostly 
associated with the slightly more attractive $^3S_1$ state in the Usmani potential. This can be easily understood by comparing 
the scattering parameters (see Table \ref{tabNloUsmani}) of the Usmani and the chiral potential. While the scattering lengths in the singlet state are almost identical, 
they do differ in the triplet state, where the Usmani potential is slightly more attractive, resulting in an enhanced correlation.
\\
In addition for the NLO correlation function we compare the exact solution to the approximate Lednick\'{y} model.
As expected we observe a significant deviation between the exact and approximate solution at small source radii. 
The Lednick\'{y} model produces an enhanced correlation signal because it contains the NLO scattering parameters that account 
only for the average attractive interaction. On the contrary the correlation function obtained feeding the exact NLO wave function in CATS 
shows the repulsive component expected within this model.
\\
It is clear that for sources $\ge 2$ fm the correlation function is not sensitive to the repulsive core both 
for p-p and p-$\mathrm{\Lambda}$ pairs. However, in the case of smaller 
sources, as shown in the last panel of Fig.~\ref{fig:res2}, the correlation functions split up and remain different even for relative momenta $k>150$ MeV.
This region is commonly used in femtoscopy analyses as a correlation free baseline to which the overall normalization of $C(k)$ is applied. 
Due to to the possible non-flat correlation in that region particular care has to be considered in including the correct potential in the correlation function.
\\~\\
In Fig.~\ref{fig:res3} the emitting source has been directly taken from the EPOS model simulating pp collisions at $\sqrt{s}=7$~TeV.
In this transport code the evaluated source peaks at very low distances and has a long tail, as already shown in Fig.~\ref{fig:EposSource}. 
Moreover the EPOS source presents a non-trivial angular dependence of the emission source. 
The sources for both the p-p and p-$\mathrm{\Lambda}$ systems in EPOS are the same. 
However, if a Gaussian parametrization is employed 
to reproduce the correlation functions obtained with EPOS, different source sizes $r_0$ for both systems are needed. 
This suggests that when investigating experimental baryon-baryon correlations in small systems, particular care 
has to be taken in the assumption of the profile of the emitting source.
\\~\\
Results for $\mathrm{K^-}$-p and $\mathrm{K^+}$-p correlations for different source sizes are shown in the upper and lower panel of Fig.~\ref{fig:res4}. 
The correlation function including only the Coulomb interaction is plotted in order to see the effect of the strong potential.
\\
The $\mathrm{K^-}$-p correlation function is deeply affected by the strong interaction already at $k\approx 200$ MeV and it 
dominates the low $k$ region. As expected both these effects are enhanced when the source gets smaller.
\\
The opposite scenario is depicted for the $\mathrm{K^+}$-p pair, where the Coulomb and the strong repulsion lead to an anti-correlated signal at low momenta.
\\~\\
In Fig.~\ref{fig:resXi} we show the expected theoretical p-$\mathrm{\Xi}^-$ correlation function. We compare $C(k)$ for a pure Coulomb interaction with the 
correlation function including the preliminary strong potential from the HAL QCD collaboration. The included channels are 
the $(I=0,~S=0)$, $(I=0,~S=1)$, $(I=1,~S=0)$ and $(I=1,~S=1)$, where only s-waves are considered. 
There is a clear enhancement of the correlation signal at small relative momenta $k$, which is a result of the attractive $I=0$ channel. 
This effect is further increased for small emitting sources.
\\~\\
In Fig.~\ref{fig:res5} we present results for $\mathrm{\Lambda}$-$\mathrm{\Lambda}$ analysis with several interaction potentials (see Sec.~\ref{interactionyy}). 
As a comparison the correlation function obtained only with the quantum statistics included is presented. 
The latter is responsible for the limit value $C(k) =0.5$ as the relative momentum $k$ tends to zero.
\\
The correlation function shows a strong sensitivity to the interaction potential and moreover the shape of $C(k)$ is deeply affected by the source size.
This behaviour can be explained easily by looking at the potentials shown 
in Fig.~\ref{fig:SourceAndPotentialsL} and their corresponding scattering parameters listed in Table~\ref{tabyy}. 
The most binding potentials, NF$42$ and ND$56$, lead to a correlation function that is significantly enhanced for smaller sources as it becomes sensitive to the deep attractive contribution. 
The weakly attractive potentials, NF$50$ and fss2, also show an enhancement compared to the quantum statistics baseline, nevertheless the correlation signal 
is much weaker due to the presence of a large repulsive core.
\\~\\
When making predictions about the correlation function it is important to take into account the expected experimental limitations. In particular the 
momentum resolution and the residual correlations contributing to the femtoscopic signal can distort the correlation function~\cite{OliThesis}. 
To estimate how the p-$\mathrm{\Xi}^-$ and $\mathrm{\Lambda}$-$\mathrm{\Lambda}$ 
correlation functions would be affected by experimental corrections, we use the linear decomposition of the correlation function as described in~\cite{OliThesis}. 
In short one assumes that only a fraction $\lambda$ of the particle pairs represent the actual correlation of interest 
and the rest of the pairs are either missidentified or stemming from the decays of resonances (feed-down). 
For the purpose of a qualitative prediction 
we will assume that all feed-down contributions have a flat correlation distribution. 
In that case the experimental correlation function $C_{exp}(k)$ is expressed by
\begin{align}\label{eq:Cexp}
C_{exp}(k)-1=\lambda(C(k)-1).
\end{align}
In Fig.~\ref{fig:LLexp} we plot the estimated experimental correlation functions for p-$\mathrm{\Xi}^-$ and 
$\mathrm{\Lambda}$-$\mathrm{\Lambda}$ pairs. For both systems we apply the momentum smearing matrix used for the p-$\mathrm{\Lambda}$ correlation in~\cite{OliThesis}. 
The $\lambda$ coefficient for the $\mathrm{\Lambda}$-$\mathrm{\Lambda}$ correlation is taken from the same reference and has the value of $32\%$.
To obtain a reasonable $\lambda$ coefficient for p-$\mathrm{\Xi}^-$ 
we use the proton feed-down and impurities as evaluated in~\cite{OliThesis}, while for the $\mathrm{\Xi}^-$ particle we assume $90\%$ purity and feed-down 
fractions of $32\%$ from $\mathrm{\Xi}(1530)$ \cite{Abelev:2014qqa} and $\approx 1\%$ from $\mathrm{\Omega}^{-}$. 
The corresponding $\lambda$ coefficient for the genuine p-$\mathrm{\Xi}^-$ correlation has a value of $52\%$.\\
One can see in Fig.~\ref{fig:LLexp} that for the $\mathrm{\Lambda}$-$\mathrm{\Lambda}$ case the effect of the 
different potentials on the correlation function is significantly damped by the inclusion of expected experimental effects. 
This implies that very large statistics are needed to reach a high precision in the determination of the scattering parameters or in testing different models.
For the p-$\mathrm{\Xi}^-$ case the effect of the strong potential is clearly distinguishable from the Coulomb-only interaction even if realistic $\lambda$ parameters are considered.
\\~\\
In order to apply the CATS framework to experimental data we present, in Fig.~\ref{fig:HADESfit}, the result for 
the p-p correlation function obtained in p-Nb collisions at $\sqrt{s_\textrm{NN}}=3.18$~GeV~\cite{Adamczewski-Musch:2016jlh}. 
We evaluated $C(k)$ in CATS using the A$v_{18}$ potential with s- and p-waves included and performed a fit 
to the data by multiplying $C(k)$ with a normalization constant $N$. The source is assumed to be Gaussian and the source size is a free fit parameter. 
The HADES analysis extracts a radius of $r_0=2.02\pm0.01(\textrm{stat})$~fm and our fit value is $2.01\pm0.01(\textrm{stat})$~fm, 
which is in perfect agreement with the published data.

\section{Conclusions and outlooks}\label{conclusion}

In this paper we presented the new femtoscopy analysis tool CATS, which allows for an exact computation of the correlation function for different sources and potentials. 
To obtain a solution valid for any source size CATS solves the Schr\"odinger equation exactly, thus providing 
the possibility to investigate not only heavy-ion reactions but small collision systems as well.
\\
In the present work we show the theoretical correlation functions for 
the p-p, p-$\mathrm{\Lambda}$, $\mathrm{K(\bar{K})}$-p, p-$\mathrm{\Xi}^-$ and $\mathrm{\Lambda}$-$\mathrm{\Lambda}$ particle pairs, 
by using different models for the interaction potential.
\\
A comparison between CATS and CRAB reveals that for an emitting source of around $2~$fm the correlation functions 
are quite similar. Nevertheless, for small colliding systems, such as pp at TeV energies, the exact treatment of the Schr\"odinger equation 
strongly deviates from the approximate solution of CRAB. 
Thus CATS provides a unique opportunity to investigate the short-range interaction between particles in small systems.
\\
We show that both the p-p and p-$\mathrm{\Lambda}$ correlation functions are affected by the repulsive core of their interaction potentials, 
which is mostly evident for small emitting sources. This effect is particularly relevant for momenta above 100~MeV, where 
the exact solution deviates significantly compared to approximate models like Lednick\'y.
\\
Moreover for the p-p and p-$\mathrm{\Lambda}$ correlations we test different emission sources by using the EPOS transport model. 
In Fig.~\ref{fig:EposSource} we show that the emitting source provided by EPOS is completely different from the traditionally Gaussian 
form used so far in femtoscopic analyses. 
The results presented here suggest the need for a more detailed analysis of the profile of the emitting source 
beyond the traditional Gaussian parametrization, whenever dealing with small systems.
\\
The investigation of the $\mathrm{K(\bar{K})}$-p correlation function shows a relevant modification due to the strong interaction. 
This effect increases with a decreasing source size.
\\
The p-$\mathrm{\Xi}^-$ correlation function is calculated by employing a preliminary local potential obtained from recent lattice HAL QCD results. 
We observe that for small emitting sources, as expected in pp collision systems, the strong interaction modifies the correlation function significantly. 
This implies that future experimental femtoscopic data on the p-$\mathrm{\Xi}^-$ correlation should be sensitive to this effect. 
This statement is still valid if one includes momentum resolution corrections and residual effects in the correlation function.
\\
For the $\mathrm{\Lambda}$-$\mathrm{\Lambda}$ correlation function we show results for different interaction potentials, based on Nijmegen boson-exchange models and quark models.
Starting from source sizes  of $r_0=2$~fm, the corresponding $C(k)$ shows a relevant sensitivity to the strong potentials. As the source radius decreases the correlation
function is modified by the attractive contribution common to all of the above-mentioned potentials. 
We also show that the inclusion of the expected experimental effects makes the separation of the different potentials extremely challenging and will 
require a lot of statistics.
\\
CATS was designed such that it can easily be used with an external fitter. To demonstrate this we perform a 
refit of the p-p correlation function measured by the HADES experiment in p-Nb reactions at $\sqrt{s_\textrm{NN}}=3.18$~GeV. 
The obtained results are in a perfect agreement with the published data.
\\
The experimental database of femtoscopy 
is constantly growing, thus in the near future it will be essential to perform global analyses over all available data in order to get 
the best possible constraints on the interaction potentials. The flexibility of CATS allows it to be used as the base of more complex frameworks 
and is therefore ready to fulfill the requirements of the upcoming experimental analyses.

\section*{Acknowledgements}
The authors are grateful to Dr. Stefano Gandolfi, Prof. Johann Haidenbauer,  Prof. Tetsuo Hatsuda, Prof. Norbert Kaiser, Prof. Kenji Morita and Prof. Wolfram Weise 
for the stimulating discussions and the useful contributions to the work presented here.\\
This work is supported by the 
Deutsche Forschungsgemeinschaft through grant SFB 1258 
"Neutrinos and Dark Matter in Astro- and Particle Physics".
We also  gratefully acknowledges the
support by the grants BMBF05P15WOFCA, DFG EClust 153, MLL, TU M\"unchen (Germany).
\appendix
\section{}\label{appA}

The ``Correlation Analysis Tool using the Schr\"odinger equation'' is implemented in a C++ class called CATS, alongside with few supporting classes 
that are provided in the CATS package, and can be found here: http://www.denseandstrange.ph.tum.de/index.php?id=78. 
In this section section we would like to discuss some of the technical details regarding the working principles of CATS. 
All information refers to CATS version 2.7, any future changes will be marked in the release notes.
\\~\\
The CATS software package contains basic examples highlighting the main input parameters that the user needs to set up in their code. 
Those include the basic properties of the particle pair, e.g. reduced mass, charges etc. We would like to focus on the two most important inputs, namely the 
emission source and the interaction potential.

\subsection{The source function}

Commonly the source function is considered to have only $r$ dependence, $r$ being the relative distance between the two particles. However 
CATS uses a more generic definition which allows   an additional dependence on the relative momentum $k$ and the scattering angle $\theta = \sphericalangle(\vec{r},\vec{k})$. 
The specific implementation of the source and its parameters in CATS allows to link it to an external framework to perform data-fitting (see Fig.~\ref{fig:HADESfit}). 
\\
Alternately the source can be inserted into CATS from an OSCAR 1997 A file \cite{OSCAR}.
This format is commonly used by transport codes in relativistic heavy ion collisions and it contains event-by-event information about 
the position and the momentum four-vectors of all particles produced in the simulation.
Note that obtaining the source from simulations leads to statistical uncertainties which result in a limited precision when determining the source function. 
Thus CATS also computes the uncertainty associated with the source function and propagates it to the correlation function. 



\subsection{Computing the wave-function}
The heart of CATS is the numerical solver for computing the wave function. The two body problem can be reduced to a one-body problem 
simply by using the reduced mass $\mu = (m_1m_2)/(m_1+m_2)$ and the reduced relative momentum $k = \frac{1}{2}|\vec{p}_1-\vec{p}_2|$, where 
$\vec{p}_{1}$ and $\vec{p}_2$ are the momenta of the two particles evaluated in their center of mass (CM) frame of reference.
\\
To obtain the SE solution we follow the standard procedure of separation of variables and partial waves decomposition (Eq.~(\ref{eq:SepVar})) 
information on which, including the standard notation, can be found in any quantum mechanics textbook (e.g.~\cite{Griffiths:QM}) 
 \begin{equation}\label{eq:SepVar}
 \begin{split}
  \psi_k(\vec{r})&=R_k(r)Y(\theta)=\sum_{l=0}^{l_{max}}R_{k,l}(r)Y_l(\theta)\\&=\sum_{l=0}^{l_{max}}i^l(2l+1)R_{k,l}(r)P_l(\textrm{cos}\theta).
 \end{split}
 \end{equation}
For low-energy problems, as the one at hand, this series 
converges quite fast. 
The Legendre polynomials $P_l$ are easily computed using the GNU scientific library (GSL) \cite{GNUGSL}, hence the only thing left to evaluate 
is the radial wave function $R_{k,l}(r)$, which satisfies the radial Schr\"odinger equation (RSE)
\begin{equation}\label{eq:RSE}
 \frac{\mathrm{d}^2u_{k,l}(r)}{\mathrm{d}r^2}=\left[\frac{2\mu V_{I,s,l,j}(r)}{\hbar^2}+\frac{l(l+1)}{r^2}-k^2\right]u_{k,l}(r),
\end{equation}
where $u_{k,l}(r)=rR_{k,l}(r)$ and $V_{I,s,l,j}(r)$ is the interaction potential between the particle species of interest. As marked by the subindices 
the potential can depend on all quantum numbers describing the pair and consequently the solutions for the radial wave functions will differ between the different states.  
How to combine the different solutions to obtain the final correlation function will be explained at the end of this subsection, but first let us turn the focus 
to solving the Schr\"odinger equation for fixed values of $k$, $I$, $s$, $l$ and $j$.
\\
To avoid the overuse of indices, the radial wave functions $u_{k,l}(r)$, $R_{k,l}(r)$ and the potential $V_{I,s,l,j}(r)$ 
will be written as $u(r)$, $R(r)$ and $V(r)$ respectively. 
In order to solve Eq.~(\ref{eq:RSE}) numerically we employed a simple Euler-method equipped with an adaptive grid which 
automatically adjusts its step size when computing $u(r)$. The basic idea is to make the step larger when computing regions 
of the wave function that have a strong non-linear behavior, i.e. a large second derivative. The discretized version of 
Eq.~(\ref{eq:RSE}) using the Euler method and a dynamic step size reads
 \begin{equation}\label{eq:ODEsol2}
  u_{i+1} = u_i\left(1+\frac{\Delta_{i}}{\Delta_{i-1}}\right)-u_{i-1}\frac{\Delta_{i}}{\Delta_{i-1}}+\underbrace{\mathcal{F}_i u_i\Delta_{i}^2}_{\approx \text{const}},
 \end{equation}
with 
\begin{equation}\label{eq:PropFun}
 \mathcal{F}_i = \left[\frac{2\mu V(r)}{\hbar^2}+\frac{l(l+1)}{r^2}-k^2\right],
\end{equation}
where $u_i$ is the value of $u(r)$ at the i-th point on the discrete $r$-grid, $\Delta_i = r_{i+1}-r_{i}$ is the distance on the grid to the next evaluation point. 
The relation $u''_{i} = \mathcal{F}_i u_r$ is given by combining Eq. \ref{eq:RSE} and Eq. \ref{eq:PropFun}, thus in order to adjust the step size 
one could simply keep the last term in Eq.~(\ref{eq:ODEsol2}) constant. This ensures that $\Delta_i^2\propto 1/u''_{i}$ and the non-linear increase 
of $u_{i+1}$ is kept constant throughout the computation.
We have explored the possibility to select a more sophisticated numerical method, e.g. a Numerov method. However the inclusion of an adaptive grid tends to become 
quite complicated for higher-order numerical solvers. We have found out that the Euler method with an adaptive provides a faster solution, 
without compromising accuracy or precision, compared to the Numerov method without an adaptive grid. 
For this reason we have decided to employ the former.
\\~\\
One additional problem that arises when solving the Schr\"odinger equation is related to the boundary conditions. In particular Eq.~(\ref{eq:ODEsol2}) 
can be used to get the full wave function only if the first two points $u_{0,1}$ are known. Note that $u(r)=rR(r)$, hence 
$u(0)=u_0=0$ can be used as the starting point of the solver. However the second point of the wave function depends on the potential and is thus a priori not known. 
Nevertheless if the second point $u_1$ is chosen randomly Eq.~(\ref{eq:ODEsol2}) will still yield the correct shape of the wave function 
at the expense of a wrong absolute normalization. To obtain the correct normalization a second boundary condition is needed.
\\ 
A straightforward approach to overcome this issue is to use the asymptotic solution. For a short range interaction potential 
$V(r)$, the evolution of the asymptotic wave function is governed by a phase shifted free or Coulomb wave. 
CATS is able to check when the solution has reached its asymptotic region and then simply match the numerical solution to the asymptotic one. 
This procedure allows to determine not only the normalization of $u(r)$ but the phase shift of the potential $V(r)$ as well (see Fig.~\ref{fig:ppphaseshift}).
The total wave function $\psi_k(\vec{r})$ is given by the sum over all $l$ partial waves
\begin{equation}\label{eq:SumPWs}
 \psi_k(\vec{r})=\sum_{l=0}^{l_{max}}\psi_{k,l}(\vec{r})=
 \underbrace{\sum_{l=0}^{l_{n}}\psi_{k,l}(\vec{r})}_{\text{numerical}}+\underbrace{\sum_{l=l_{n}+1}^{l_{max}}\psi_{k,l}(\vec{r})}_{\text{asymptotic}}.
\end{equation}
Here the value of $l_{max}$ is determined by the condition for convergence, namely when $l>l_{max}$ the partial waves are practically zero in the asymptotic range. 
The sum is than split in two parts - first all numerically evaluated partial waves are summed up and 
than all remaining partial waves are added using their asymptotic solution obtained from GSL. 
\\~\\
Another important thing to consider is the possibility to have different spin and isospin states.
Moreover since the total angular momentum $J$ is a good quantum number for potentials with spin-orbit terms, the degeneracy in $J$ has to be taken into account as well.
All of these contributions will result in different partial wave states $^{2S+1} L_J$  
and the user should provide CATS with all relevant information in order to get a correct total correlation function.  
The way this is achieved is by introducing the notation of interaction channels (see Fig.~\ref{fig:Channels}). 
An interaction channel is defined as the total wave function for a specific
spin and isospin state and each partial wave that is included in the computation is considered to be in a fixed $J$ state. The user should carefully examine all possible permutations 
to include each channel that contributes to the final correlation and should provide CATS with the probabilities that the 
particle pair can be found in particular channel. 
\\
We provide an example for the p-p interaction. For this system the isospin can only be 1, hence it is 
not considered in the calculation. However the total spin $S$ of 
the system can be either 0 or 1. Since those states correspond to a singlet and a triplet configuration respectively, as long as there is no preferred polarization the probability to be 
in the singlet state is $1/4$ and in the triplet $3/4$. Those weights can be computed from the simple formula regarding spin degeneracy
\begin{equation}\label{eq:SpinDeg}
 w_{(S)}=\dfrac{(2S+1)}{(2s_1+1)(2s_2+1)},
\end{equation}
where $s_1$ and $s_2$ are the spins of the individual particles. If one would consider only the $s$-waves in the computation the relevant partial waves are $^1S_0$ and $^3S_1$. 
In fact for p-p, due to the quantum statistics of identical fermions, the $^3S_1$ state is not allowed
and therefore no potential will be considered in CATS. 
Nevertheless it is very important to note that the weights of $1/4$ and $3/4$ still need to be set, since 
even if the $^3S_1$ partial wave is canceled out for p-p, the total wave function of the $S=1$ is still included in $C(k)$ with a flat contribution. 
The situation becomes a bit more complex if the $p$-waves are to be considered as well. 
In the $S=0$ the $^1P_1$ is the only $p$-partial wave and for p-p it is canceled out again due to the quantum statistics. However the $S=1$ state has 3 possible 
$p$-waves, namely $^3P_0$, $^3P_1$ and $^3P_2$. These states have the same $S$ and $L$ quantum numbers, but different $J$. Hence to compute their relative contributions one 
needs to consider the degeneracy in $J$ which is given by
\begin{equation}\label{eq:JDeg}
 w_{(J)}=\dfrac{(2J+1)}{(2L+1)(2S+1)}.
\end{equation}
The total weights will be given by:
\begin{align}\label{eq:FinalWeights}
 &w_{(S,L,J)}=w_{(S)}\cdot w_{(J)}=\\
 &\dfrac{(2S+1)}{(2s_1+1)(2s_2+1)}\dfrac{(2J+1)}{(2L+1)(2S+1)}.
\end{align}
In Fig.~\ref{fig:Channels} we show a schematic representation of the above example.\\
From the user's perspective, the inputs that CATS has to be provided with are the number of interaction channels, the number of partial waves to be included in each channel and 
finally the potentials for each partial wave state. 
In the example from 
Fig.~\ref{fig:Channels} the user needs to define 4 channels each containing two partial waves. However, due to the Pauli blocking, the 
$^1P_1$ and $^3S_1$ states can be safely left undefined in the code.

\bibliography{biblio}{}

\begin{thebibliography}{10}
\providecommand{\url}[1]{{#1}}
\providecommand{\urlprefix}{URL }
\expandafter\ifx\csname urlstyle\endcsname\relax
  \providecommand{\doi}[1]{DOI \discretionary{}{}{}#1}\else
  \providecommand{\doi}{DOI \discretionary{}{}{}\begingroup
  \urlstyle{rm}\Url}\fi

\bibitem{Pratt:1984su}
S.~Pratt, Phys. Rev. Lett. \textbf{53}, 1219 (1984).
\newblock \doi{10.1103/PhysRevLett.53.1219}

\bibitem{Lisa:2005dd}
M.A. Lisa, S.~Pratt, R.~Soltz, U.~Wiedemann, Ann. Rev. Nucl. Part. Sci.
  \textbf{55}, 357 (2005).
\newblock \doi{10.1146/annurev.nucl.55.090704.151533}

\bibitem{Agakishiev:2011zz}
G.~Agakishiev, et~al., Eur. Phys. J. \textbf{A47}, 63 (2011).
\newblock \doi{10.1140/epja/i2011-11063-x}

\bibitem{Kotte:2004yv}
R.~Kotte, et~al., Eur. J. Phys. \textbf{A23}, 271 (2005).
\newblock \doi{10.1140/epja/i2004-10075-y}

\bibitem{Aggarwal:2007aa}
M.M. Aggarwal, et~al.,   (2007)

\bibitem{Adams:2004yc}
J.~Adams, et~al., Phys. Rev. C \textbf{71}, 044906 (2005).
\newblock \doi{10.1103/PhysRevC.71.044906}

\bibitem{Aamodt:2011mr}
K.~Aamodt, et~al., Phys. Lett. \textbf{B696}, 328 (2011).
\newblock \doi{10.1016/j.physletb.2010.12.053}

\bibitem{Adams:2005ws}
J.~Adams, et~al., Phys. Rev. \textbf{C74}, 064906 (2006).
\newblock \doi{10.1103/PhysRevC.74.064906}

\bibitem{Anticic:2011ja}
T.~Anticic, et~al., Phys. Rev. \textbf{C83}, 054906 (2011).
\newblock \doi{10.1103/PhysRevC.83.054906}

\bibitem{Chung:2002vk}
P.~Chung, et~al., Phys. Rev. Lett. \textbf{91}, 162301 (2003).
\newblock \doi{10.1103/PhysRevLett.91.162301}

\bibitem{Agakishiev:2010qe}
G.~Agakishiev, et~al., Phys. Rev. \textbf{C82}, 021901 (2010).
\newblock \doi{10.1103/PhysRevC.82.021901}

\bibitem{Adamczyk:2014vca}
L.~Adamczyk, et~al., Phys. Rev. Lett. \textbf{114}(2), 022301 (2015).
\newblock \doi{10.1103/PhysRevLett.114.022301}

\bibitem{Adam:2015vja}
J.~Adam, et~al., Phys. Rev. \textbf{C92}(5), 054908 (2015).
\newblock \doi{10.1103/PhysRevC.92.054908}

\bibitem{PhysRevD.84.112004}
K.e.a. Aamodt, Phys. Rev. D \textbf{84}, 112004 (2011).
\newblock \doi{10.1103/PhysRevD.84.112004}

\bibitem{Abelev:2012sq}
B.~Abelev, et~al., Phys. Rev. \textbf{D87}(5), 052016 (2013).
\newblock \doi{10.1103/PhysRevD.87.052016}

\bibitem{PhysRevC.91.034906}
J.e.a. Adam, Phys. Rev. C \textbf{91}, 034906 (2015).
\newblock \doi{10.1103/PhysRevC.91.034906}

\bibitem{Adam:2016iwf}
J.~Adam, et~al., Eur. Phys. J. \textbf{C77}(8), 569 (2017).
\newblock \doi{10.1140/epjc/s10052-017-5129-6}

\bibitem{PhysRevC.92.034906}
T.~Pierog, I.~Karpenko, J.M. Katzy, E.~Yatsenko, K.~Werner, Phys. Rev. C
  \textbf{92}, 034906 (2015).
\newblock \doi{10.1103/PhysRevC.92.034906}

\bibitem{crab}
S.Pratt, http://web.pa.msu.edu/people/pratts/freecodes/ crab/home.html

\bibitem{ourpaper}
In preparation

\bibitem{Koonin:1977fh}
S.E. Koonin, Phys. Lett. \textbf{70B}, 43 (1977).
\newblock \doi{10.1016/0370-2693(77)90340-9}

\bibitem{Reid:1968sq}
R.V. Reid, Jr., Annals Phys. \textbf{50}, 411 (1968).
\newblock \doi{10.1016/0003-4916(68)90126-7}

\bibitem{Lednicky:1981su}
R.~Lednicky, V.L. Lyuboshits, Sov. J. Nucl. Phys. \textbf{35}, 770 (1982).
\newblock [Yad. Fiz.35,1316(1981)]

\bibitem{Stoks:1993tb}
V.G.J. Stoks, R.A.M. Klomp, M.C.M. Rentmeester, J.J. de~Swart, Phys. Rev.
  \textbf{C48}, 792 (1993).
\newblock \doi{10.1103/PhysRevC.48.792}

\bibitem{Machleidt:1987hj}
R.~Machleidt, K.~Holinde, C.~Elster, Phys. Rept. \textbf{149}, 1 (1987).
\newblock \doi{10.1016/S0370-1573(87)80002-9}

\bibitem{Machleidt:1989tm}
R.~Machleidt, Adv. Nucl. Phys. \textbf{19}, 189 (1989)

\bibitem{Lagaris:1981mm}
I.E. Lagaris, V.R. Pandharipande, Nucl. Phys. \textbf{A359}, 331 (1981).
\newblock \doi{10.1016/0375-9474(81)90240-2}

\bibitem{Lacombe:1980dr}
M.~Lacombe, B.~Loiseau, J.M. Richard, R.~Vinh~Mau, J.~Cote, P.~Pires,
  R.~De~Tourreil, Phys. Rev. \textbf{C21}, 861 (1980).
\newblock \doi{10.1103/PhysRevC.21.861}

\bibitem{Nagels:1977ze}
M.M. Nagels, T.A. Rijken, J.J. de~Swart, Phys. Rev. \textbf{D17}, 768 (1978).
\newblock \doi{10.1103/PhysRevD.17.768}

\bibitem{Wiringa:1994wb}
R.B. Wiringa, V.G.J. Stoks, R.~Schiavilla, Phys. Rev. \textbf{C51}, 38 (1995).
\newblock \doi{10.1103/PhysRevC.51.38}

\bibitem{Entem:2015xwa}
D.R. Entem, N.~Kaiser, R.~Machleidt, Y.~Nosyk, Phys. Rev. \textbf{C92}(6),
  064001 (2015).
\newblock \doi{10.1103/PhysRevC.92.064001}

\bibitem{Epelbaum:2014sza}
E.~Epelbaum, H.~Krebs, U.G. Meißner, Phys. Rev. Lett. \textbf{115}(12), 122301
  (2015).
\newblock \doi{10.1103/PhysRevLett.115.122301}

\bibitem{Entem:2003ft}
D.R. Entem, R.~Machleidt, Phys. Rev. \textbf{C68}, 041001 (2003).
\newblock \doi{10.1103/PhysRevC.68.041001}

\bibitem{Epelbaum:1998ka}
E.~Epelbaum, W.~Gloeckle, U.G. Meissner, Nucl. Phys. \textbf{A637}, 107 (1998).
\newblock \doi{10.1016/S0375-9474(98)00220-6}

\bibitem{Kaiser:1997mw}
N.~Kaiser, R.~Brockmann, W.~Weise, Nucl. Phys. \textbf{A625}, 758 (1997).
\newblock \doi{10.1016/S0375-9474(97)00586-1}

\bibitem{Nagels:2014qqa}
M.M. Nagels, T.A. Rijken, Y.~Yamamoto,   (2014)

\bibitem{PhysRev.173.1452}
G.~Alexander, U.~Karshon, A.~Shapira, G.~Yekutieli, R.~Engelmann, H.~Filthuth,
  W.~Lughofer, Phys. Rev. \textbf{173}, 1452 (1968).
\newblock \doi{10.1103/PhysRev.173.1452}

\bibitem{PhysRev.175.1735}
B.~Sechi-Zorn, B.~Kehoe, J.~Twitty, R.A. Burnstein, Phys. Rev. \textbf{175},
  1735 (1968).
\newblock \doi{10.1103/PhysRev.175.1735}

\bibitem{pap54}
R.~Engelmann, et~al., Phys. Lett. \textbf{21}, 587 (1966)

\bibitem{Eisele:1971mk}
F.~Eisele, H.~Filthuth, W.~Foehlisch, V.~Hepp, G.~Zech, Phys. Lett.
  \textbf{37B}, 204 (1971).
\newblock \doi{10.1016/0370-2693(71)90053-0}

\bibitem{Hepp:1968zza}
V.~Hepp, H.~Schleich, Z. Phys. \textbf{214}, 71 (1968).
\newblock \doi{10.1007/BF01380085}

\bibitem{pap57}
D.~Stephen, Ph.D. thesis, University of Massachusetts p. unpublished (1970)

\bibitem{pap58}
J.~De~Swart, C.~Dullemond, Annals Phys. \textbf{19}, 485 (1962)

\bibitem{Bart:1999uh}
S.~Bart, et~al., Phys. Rev. Lett. \textbf{83}, 5238 (1999).
\newblock \doi{10.1103/PhysRevLett.83.5238}

\bibitem{PhysRevC.29.684}
A.R. Bodmer, Q.N. Usmani, J.~Carlson, Phys. Rev. C \textbf{29}, 684 (1984).
\newblock \doi{10.1103/PhysRevC.29.684}

\bibitem{Haidenbauer:2013oca}
J.~Haidenbauer, S.~Petschauer, N.~Kaiser, U.G. Meissner, A.~Nogga, W.~Weise,
  Nucl. Phys. \textbf{A915}, 24 (2013).
\newblock \doi{10.1016/j.nuclphysa.2013.06.008}

\bibitem{Gal:2016boi}
A.~Gal, E.V. Hungerford, D.J. Millener, Rev. Mod. Phys. \textbf{88}(3), 035004
  (2016).
\newblock \doi{10.1103/RevModPhys.88.035004}

\bibitem{Nakazawa:2015joa}
K.~Nakazawa, et~al., PTEP \textbf{2015}(3), 033D02 (2015).
\newblock \doi{10.1093/ptep/ptv008}

\bibitem{Nagae:2017slp}
T.~Nagae, et~al., PoS \textbf{INPC2016}, 038 (2017)

\bibitem{Hatsuda:2017uxk}
T.~Hatsuda, K.~Morita, A.~Ohnishi, K.~Sasaki, Nucl. Phys. \textbf{A967}, 856
  (2017).
\newblock \doi{10.1016/j.nuclphysa.2017.04.041}

\bibitem{Sasaki:2017ysy}
K.~Sasaki, et~al., PoS \textbf{LATTICE2016}, 116 (2017)

\bibitem{katz}
P.A.K. et~al., Phys. Rev. \textbf{D1}, 1267 (1970)

\bibitem{PhysRevLett.2.425}
R.H. Dalitz, S.F. Tuan, Phys. Rev. Lett. \textbf{2}, 425 (1959).
\newblock \doi{10.1103/PhysRevLett.2.425}

\bibitem{Chao:1973sa}
Y.A. Chao, R.W. Kraemer, D.W. Thomas, B.R. Martin, Nucl. Phys. \textbf{B56}, 46
  (1973).
\newblock \doi{10.1016/0550-3213(73)90218-6}

\bibitem{Engler:1965zz}
A.~Engler, H.E. Fisk, R.w. Kraemer, C.M. Meltzer, J.B. Westgard, Phys. Rev.
  Lett. \textbf{15}, 224 (1965).
\newblock \doi{10.1103/PhysRevLett.15.224}

\bibitem{Agakishiev:2012xk}
G.~Agakishiev, et~al., Phys. Rev. \textbf{C87}, 025201 (2013).
\newblock \doi{10.1103/PhysRevC.87.025201}

\bibitem{Moriya:2009mx}
K.~Moriya, R.~Schumacher, Nucl. Phys. \textbf{A835}, 325 (2010).
\newblock \doi{10.1016/j.nuclphysa.2010.01.210}

\bibitem{Bazzi:2011zj}
M.~Bazzi, et~al., Phys. Lett. \textbf{B704}, 113 (2011).
\newblock \doi{10.1016/j.physletb.2011.09.011}

\bibitem{Agnello:2005qj}
M.~Agnello, et~al., Phys. Rev. Lett. \textbf{94}, 212303 (2005).
\newblock \doi{10.1103/PhysRevLett.94.212303}

\bibitem{MuellerGroeling:1990cw}
A.~Mueller-Groeling, K.~Holinde, J.~Speth, Nucl. Phys. \textbf{A513}, 557
  (1990).
\newblock \doi{10.1016/0375-9474(90)90398-6}

\bibitem{Buettgen:1990yw}
R.~Buettgen, K.~Holinde, A.~Mueller-Groeling, J.~Speth, P.~Wyborny, Nucl. Phys.
  \textbf{A506}, 586 (1990).
\newblock \doi{10.1016/0375-9474(90)90205-Z}

\bibitem{Hadjimichef:2002xe}
D.~Hadjimichef, J.~Haidenbauer, G.~Krein, Phys. Rev. \textbf{C66}, 055214
  (2002).
\newblock \doi{10.1103/PhysRevC.66.055214}

\bibitem{Bender:1983cw}
I.~Bender, H.G. Dosch, H.J. Pirner, H.G. Kruse, Nucl. Phys. \textbf{A414}, 359
  (1984).
\newblock \doi{10.1016/0375-9474(84)90608-0}

\bibitem{SilvestreBrac:1997jw}
B.~Silvestre-Brac, J.~Labarsouque, J.~Leandri, Nucl. Phys. \textbf{A613}, 342
  (1997).
\newblock \doi{10.1016/S0375-9474(96)00424-1}

\bibitem{Lemaire:2002wh}
S.~Lemaire, J.~Labarsouque, B.~Silvestre-Brac, Nucl. Phys. \textbf{A700}, 330
  (2002).
\newblock \doi{10.1016/S0375-9474(01)01304-5}

\bibitem{Ikeda:2012au}
Y.~Ikeda, T.~Hyodo, W.~Weise, Nucl. Phys. \textbf{A881}, 98 (2012).
\newblock \doi{10.1016/j.nuclphysa.2012.01.029}

\bibitem{HOLZENKAMP1989485}
B.~Holzenkamp, K.~Holinde, J.~Speth, Nuclear Physics A \textbf{500}(3), 485
  (1989).
\newblock \doi{https://doi.org/10.1016/0375-9474(89)90223-6}

\bibitem{Takahashi:2001nm}
H.~Takahashi, et~al., Phys. Rev. Lett. \textbf{87}, 212502 (2001).
\newblock \doi{10.1103/PhysRevLett.87.212502}

\bibitem{Ahn:2013poa}
J.K. Ahn, et~al., Phys. Rev. \textbf{C88}(1), 014003 (2013).
\newblock \doi{10.1103/PhysRevC.88.014003}

\bibitem{Morita:2014kza}
K.~Morita, T.~Furumoto, A.~Ohnishi, Phys. Rev. \textbf{C91}(2), 024916 (2015).
\newblock \doi{10.1103/PhysRevC.91.024916}

\bibitem{Nagels:1976xq}
M.M. Nagels, T.A. Rijken, J.J. de~Swart, Phys. Rev. \textbf{D15}, 2547 (1977).
\newblock \doi{10.1103/PhysRevD.15.2547}

\bibitem{Nagels:1978sc}
M.M. Nagels, T.A. Rijken, J.J. de~Swart, Phys. Rev. \textbf{D20}, 1633 (1979).
\newblock \doi{10.1103/PhysRevD.20.1633}

\bibitem{Fujiwara:2006yh}
Y.~Fujiwara, Y.~Suzuki, C.~Nakamoto, Prog. Part. Nucl. Phys. \textbf{58}, 439
  (2007).
\newblock \doi{10.1016/j.ppnp.2006.08.001}

\bibitem{Bass:1998ca}
S.A. Bass, et~al., Prog. Part. Nucl. Phys. \textbf{41}, 255 (1998).
\newblock \doi{10.1016/S0146-6410(98)00058-1}.
\newblock [Prog. Part. Nucl. Phys.41,225(1998)]

\bibitem{Adamczewski-Musch:2016jlh}
J.~Adamczewski-Musch, et~al., Phys. Rev. \textbf{C94}(2), 025201 (2016).
\newblock \doi{10.1103/PhysRevC.94.025201}

\bibitem{OliThesis}
O.~Arnold, Study of the hyperon-nucleon interaction via femtoscopy in
  elementary systems with hades and alice.
\newblock Ph.D. thesis (2017)

\bibitem{Abelev:2014qqa}
B.B. Abelev, et~al., Eur. Phys. J. \textbf{C75}(1), 1 (2015).
\newblock \doi{10.1140/epjc/s10052-014-3191-x}

\bibitem{OSCAR}
{Bass, Pang and Werner},   (Accessed 21 February 2018).
\newblock
  \urlprefix\url{https://karman.physics.purdue.edu/OSCAR/index.php/Main_Page}

\bibitem{Griffiths:QM}
D.~Griffiths, \emph{{Introduction to Quantum Mechanics}} (Pearson Prentice
  Hall, 2004)

\bibitem{GNUGSL}
M.~Galassi, \emph{{GNU Scientific Library: Reference Manual}} (Network Theory,
  2009)

\end{thebibliography}
\bibliographystyle{spphys}

\begin{center}
 
\begin{figure*}[b!] 
 \subfigure{%
    \includegraphics[width=0.45\textwidth]{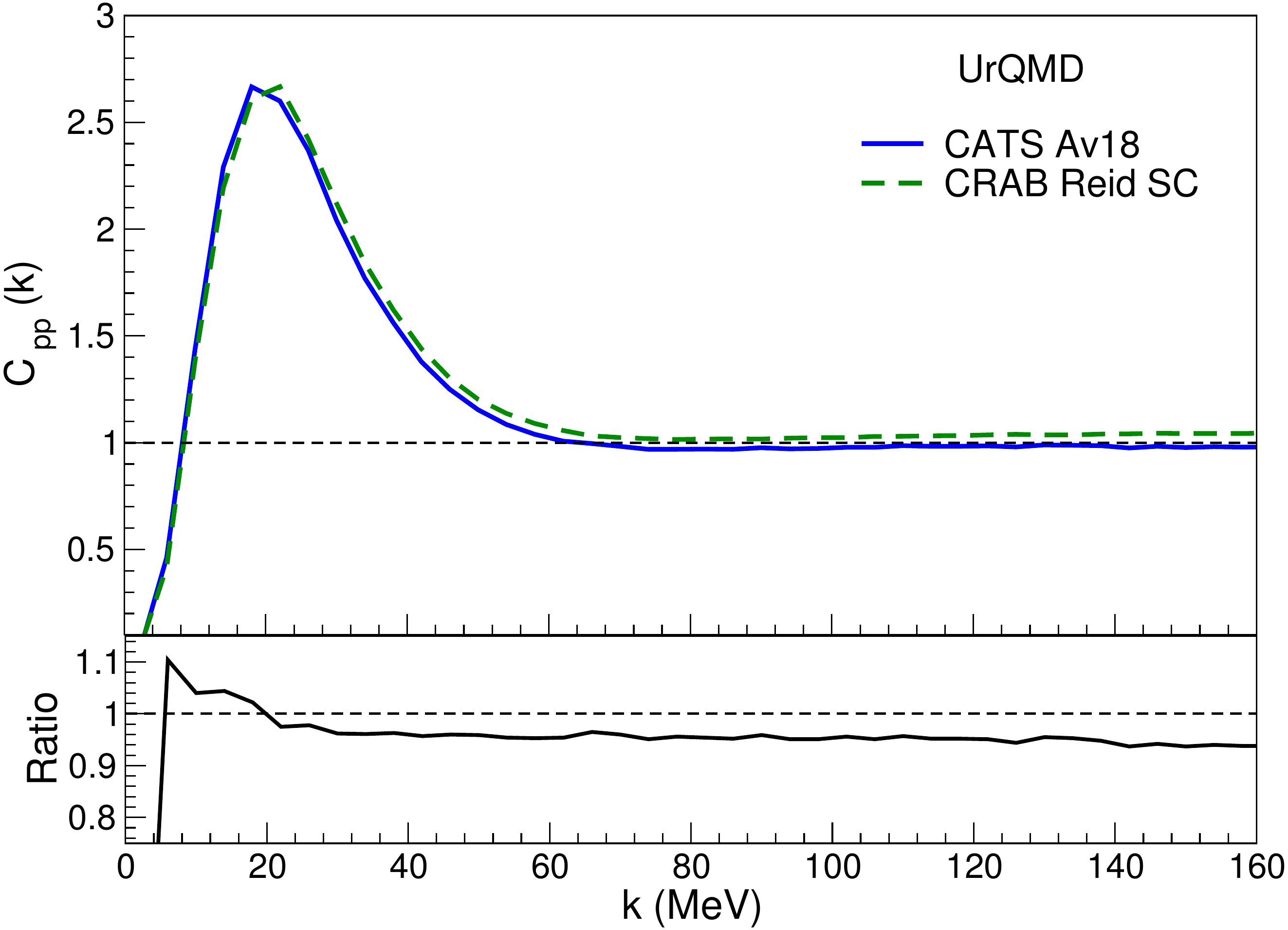} 
  } 
  \quad 
  \subfigure{%
   \includegraphics[width=0.45\textwidth]{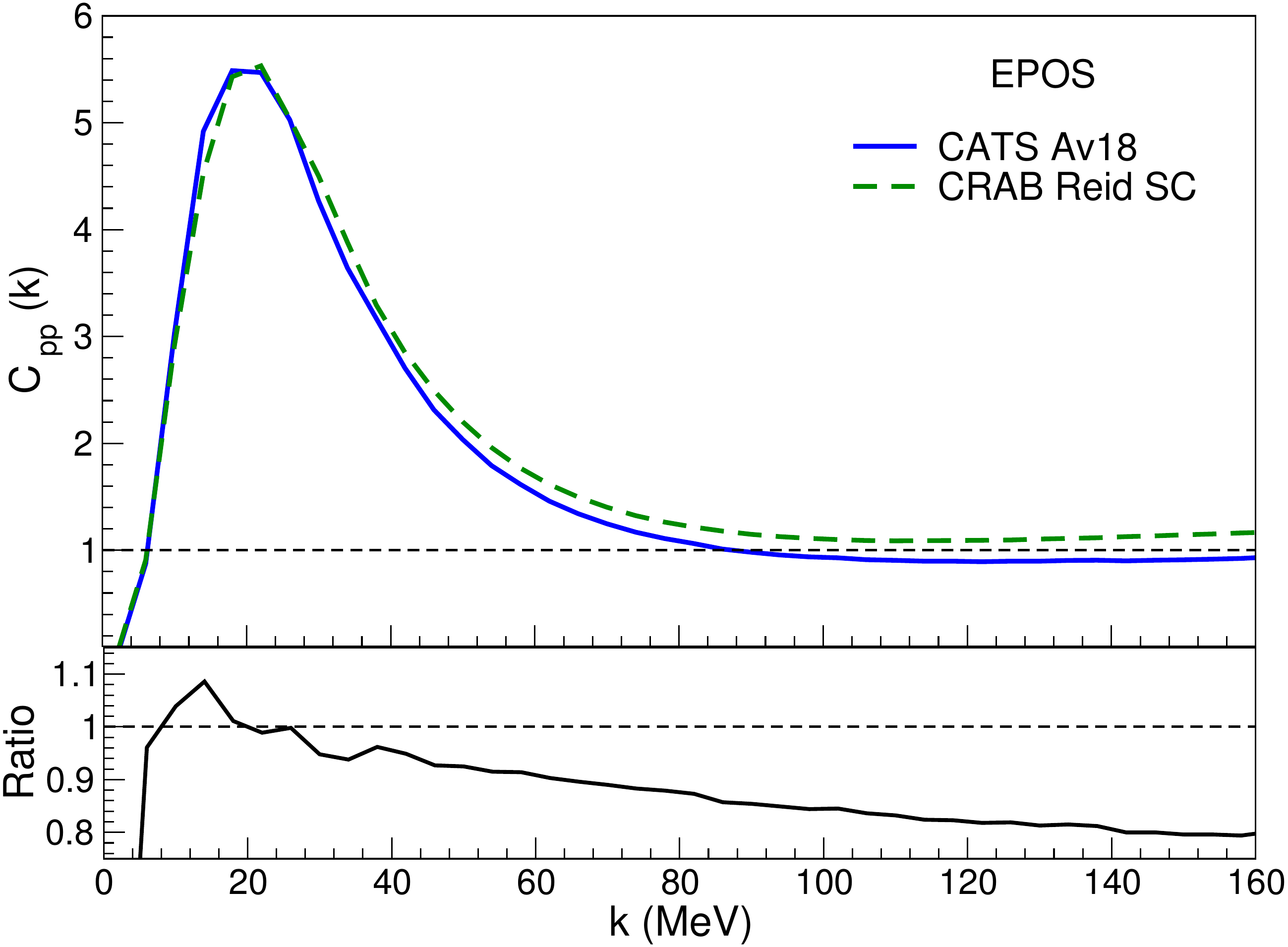} 
  } 
\caption{Correlation function for p-p pairs obtained within UrQMD simulations for p-Nb reaction at $\sqrt{s_\textrm{NN}}=3.18$ GeV (upper panel) 
and within EPOS for pp collisions at $\sqrt{s}=7$ TeV. For both setups results are presented for CATS calculations with A$v_{18}$ and CRAB 
with a modified Reid SC potential (see~\ref{interaction} for details).}\label{fig:crab1}
\end{figure*}
\begin{figure*}[tbp]
\centering{
\includegraphics[width=0.9\textwidth , scale=0.6]{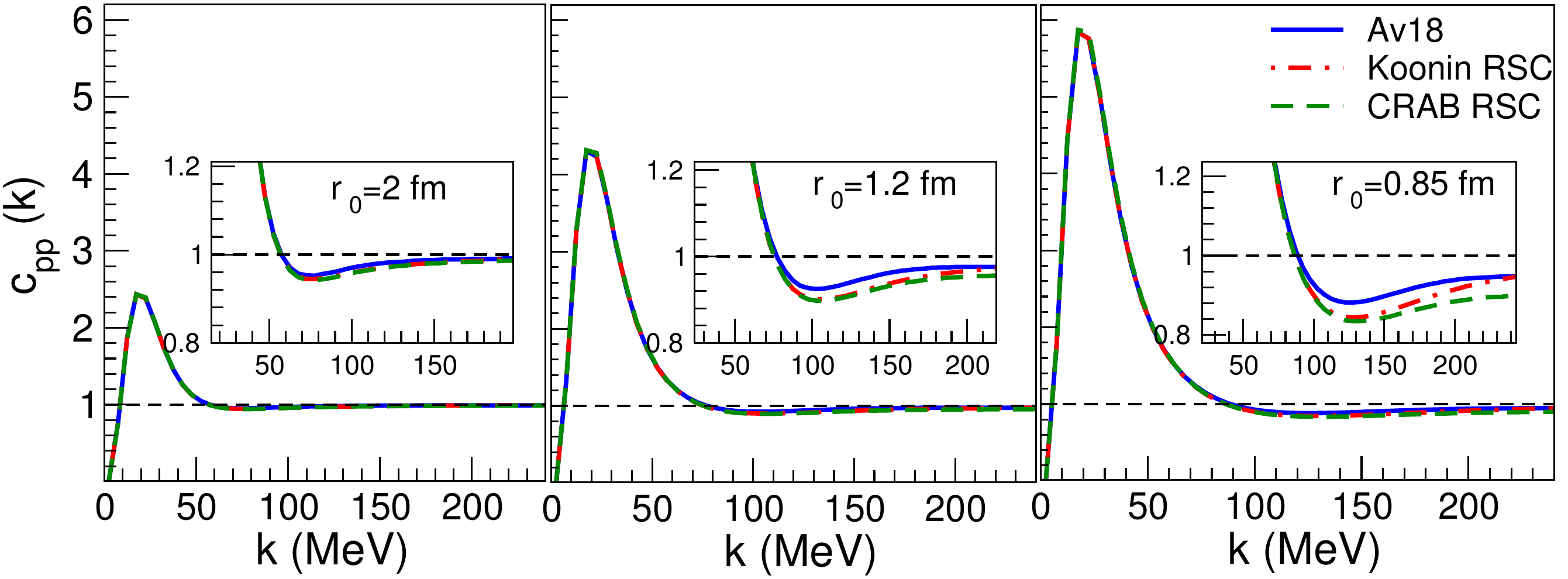}
\includegraphics[width=0.9\textwidth , scale=0.6]{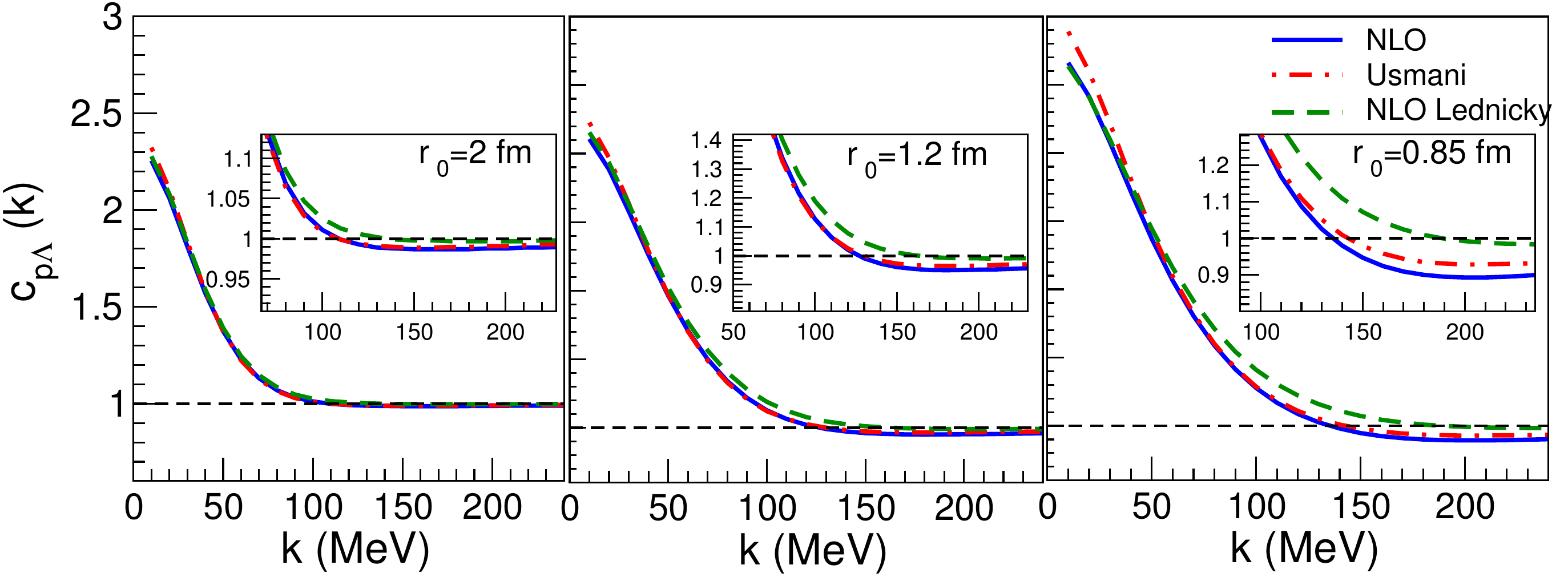}
}
\caption{Correlation functions for p-p pairs (upper panel) and p-$\mathrm{\Lambda}$ pairs (lower panel) for 
different interaction potentials (see~\ref{interaction}) and emitting source sizes.}
\label{fig:res2}
\end{figure*}
\begin{figure*}[h]
\centering{
\includegraphics[width=0.9\textwidth , scale=0.4]{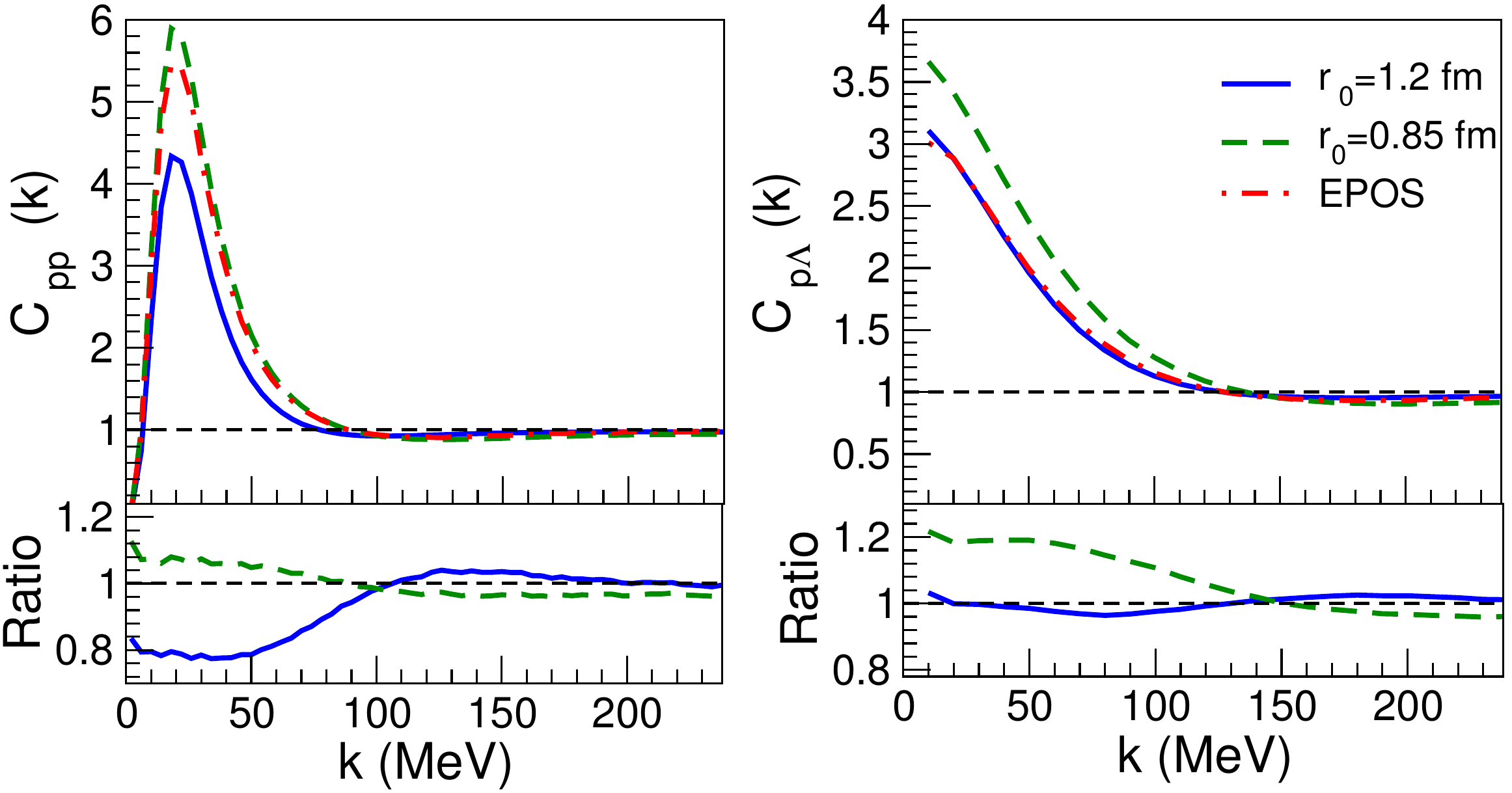}
}
\caption{Correlation functions for p-p pairs (left) with $AV_{18}$ potential and p-$\mathrm{\Lambda}$ pairs (right) with $\chi$EFT $NLO$ potential (see~\ref{interaction}). 
The calculations are performed by assuming Gaussian emitting sources of radii $r_0=1.2$ fm (solid blue line), $r_0=0.85$ fm (dashed green line) and by using 
EPOS transport model (red dot-dashed line).}
\label{fig:res3}
\end{figure*}
\begin{figure*}[tbp]
\centering{
\includegraphics[width=0.9\textwidth , scale=0.6]{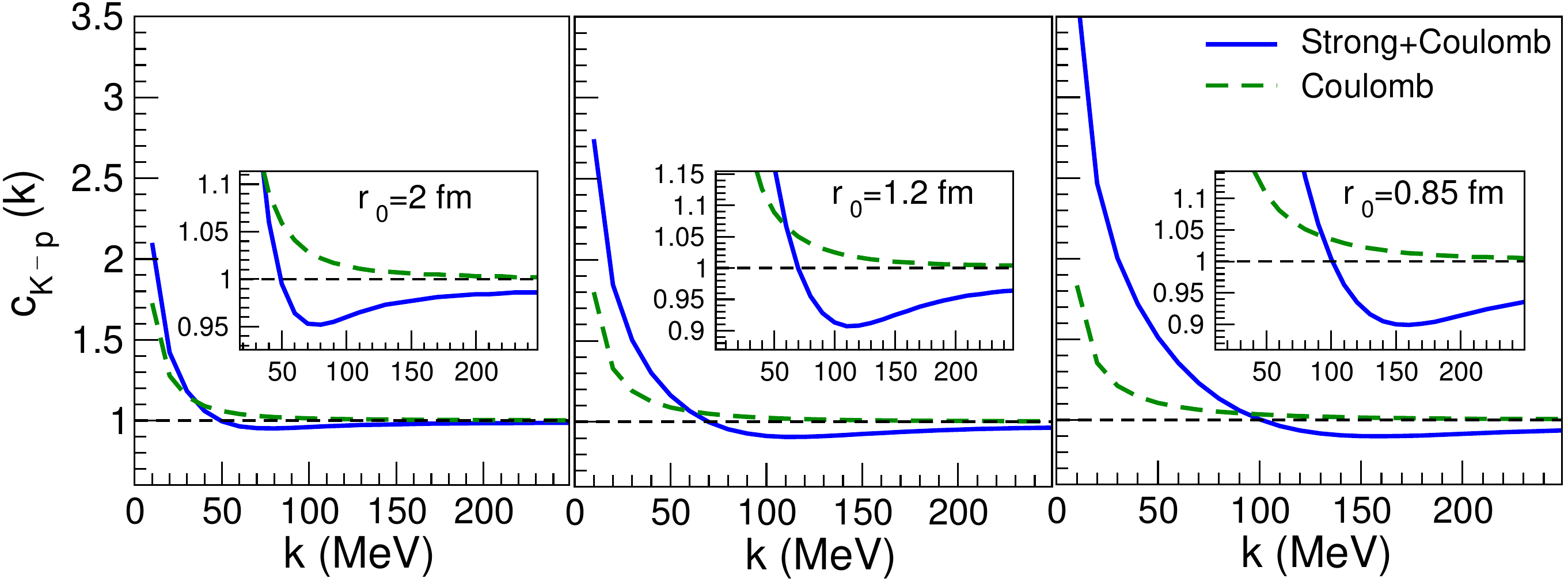}
\includegraphics[width=0.9\textwidth , scale=0.6]{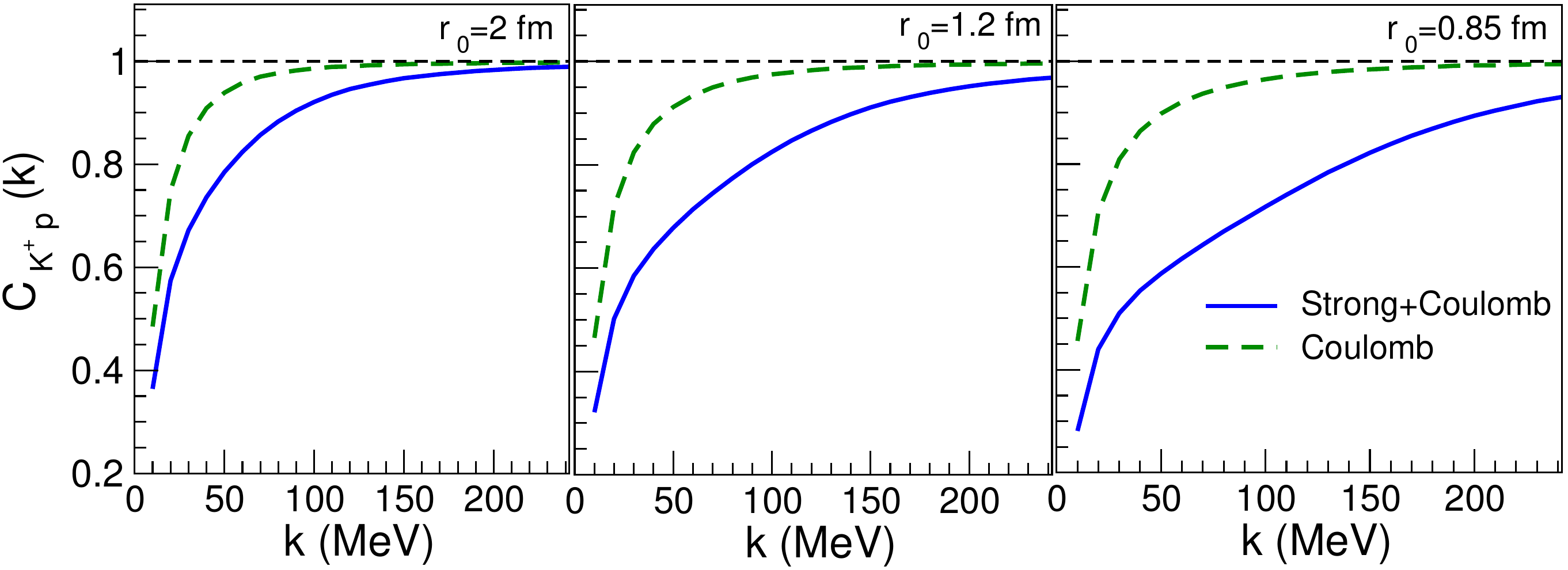}
}
\caption{Correlation functions for $\mathrm{K^-}$-p pairs (upper panel) and $\mathrm{K^+}$-p pairs (lower panel) for different emitting source sizes. 
The dashed line is the result with only the Coulomb interaction considered while the solid line shows the effect of both Coulomb and strong 
interaction based on J\"ulich model potential.}
\label{fig:res4}
\end{figure*}
\begin{figure*}[h]
\centering{
\includegraphics[width=0.75\textwidth , scale=0.6]{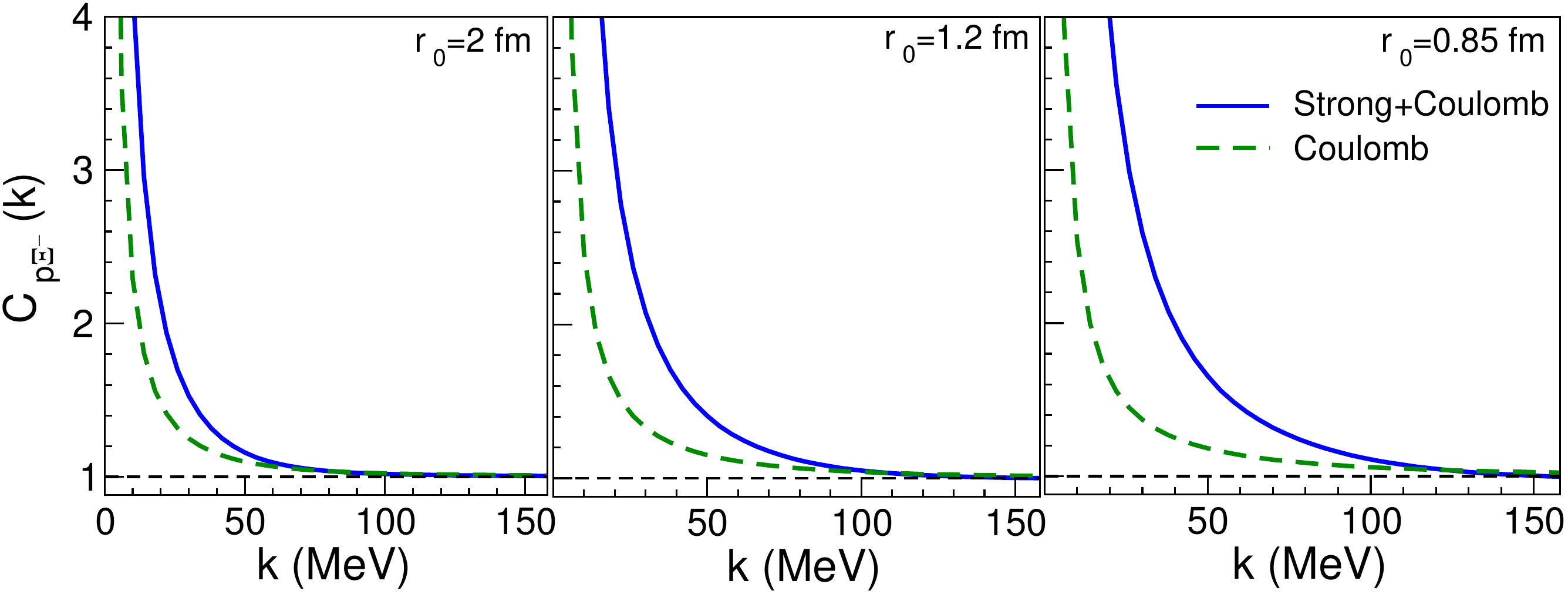}
}
\caption{Correlation functions for p-$\mathrm{\Xi}^-$ pairs for the preliminary HAL QCD potential~\cite{Sasaki:2017ysy,Hatsuda:2017uxk} with both 
$I=0$ and $I=1$ contributions (see~\ref{interaction}). Different emitting source sizes are considered.}
\label{fig:resXi}
\end{figure*}
\begin{figure*}[tbp]
\centering{
\includegraphics[width=0.75\textwidth , scale=0.8]{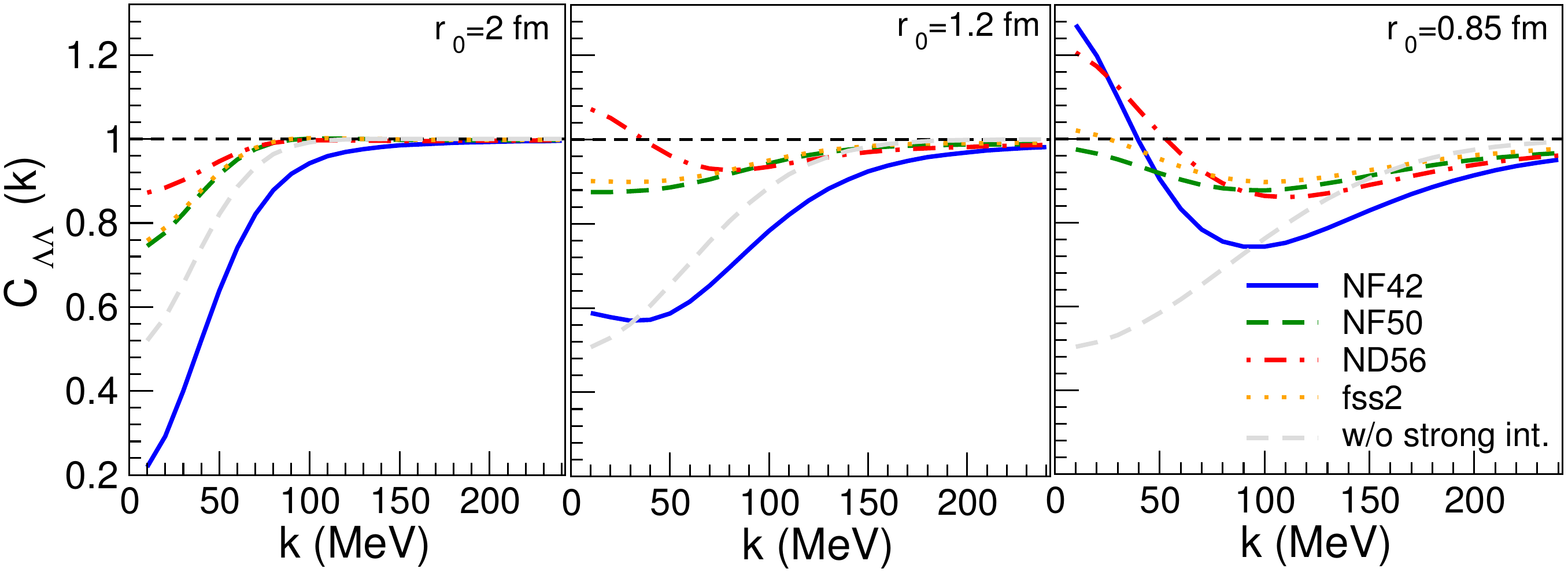}
}
\caption{Correlation functions for $\mathrm{\Lambda}$-$\mathrm{\Lambda}$ pairs for different emitting source sizes and interaction potentials. 
The dashed line is the result with only the quantum statistics considered while the solid line shows the effect of the strong potential based on 
several models (see the text for details).}
\label{fig:res5}
\end{figure*}
\begin{figure*}[tbp]
\centering{
\includegraphics[width=0.75\textwidth , scale=0.6]{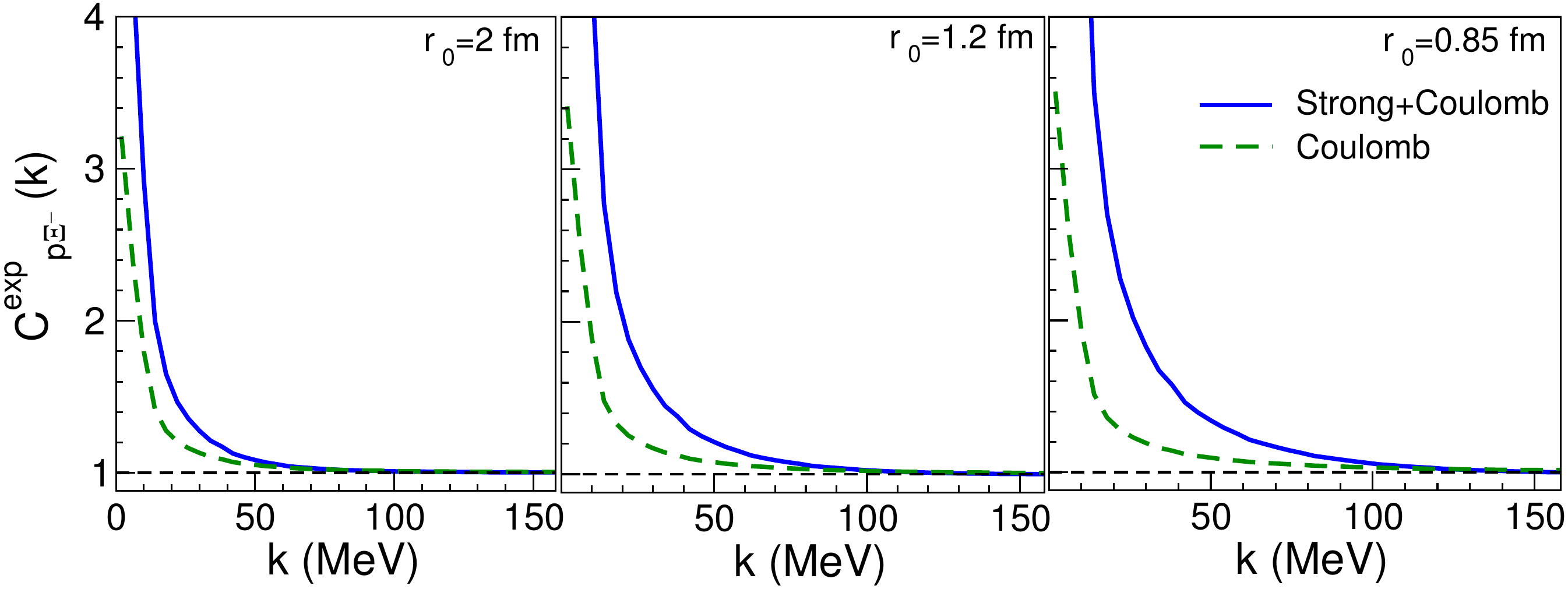}
\includegraphics[width=0.75\textwidth , scale=0.6]{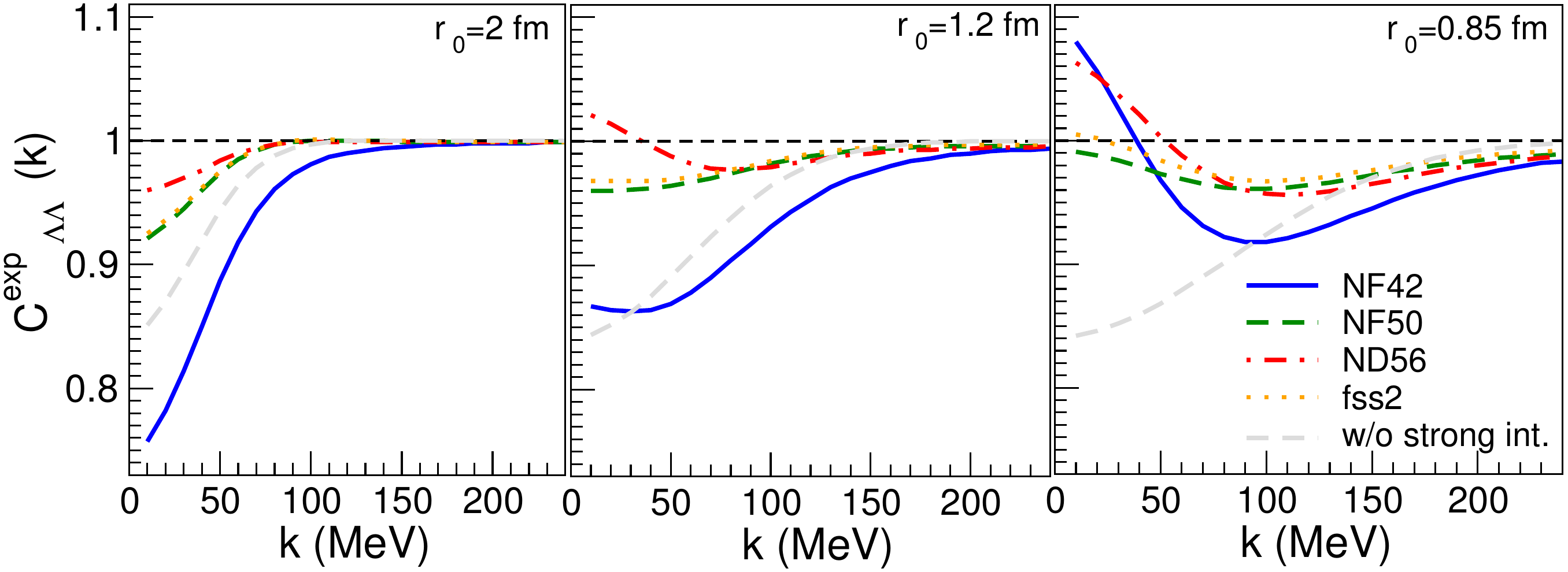}
}
\caption
{Correlation function for p-$\mathrm{\Xi}^-$ (top) and $\mathrm{\Lambda}$-$\mathrm{\Lambda}$ (bottom) with feed-down coefficients and moment 
resolution included.For p-$\mathrm{\Xi}^-$ pair, 
the preliminary HAL QCD potential~\cite{Sasaki:2017ysy,Hatsuda:2017uxk} with both $I=0$ and $I=1$ contributions has been employed (see~\ref{interaction}). 
Different emitting source sizes are considered.
}
\label{fig:LLexp}
\end{figure*}
\begin{figure*}[h]
\centering{
\includegraphics[width=0.75\textwidth , scale=0.6]{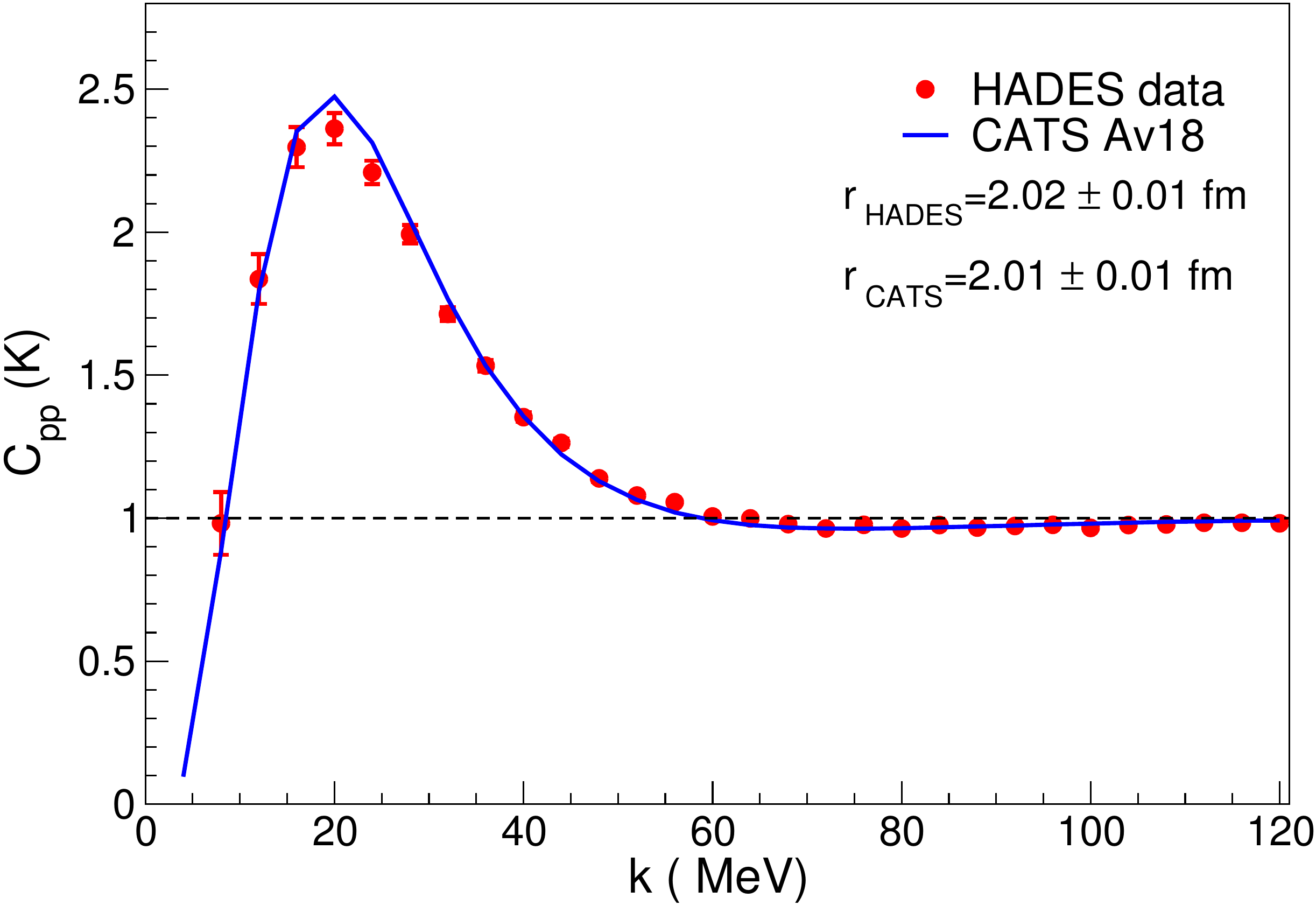}
}
\caption
{Correlation function for p-p in pNb collisions at $\sqrt{s_\textrm{NN}}=3.18$~GeV in HADES. As this fit is performed as a simple cross check only the statistical 
errors are considered. The solid line represents the results obtained within CATS by fitting the correlation function to the experimental data, introducing a normalization factor $N$.
}
\label{fig:HADESfit}
\end{figure*}
\begin{figure*}[h!]
\centering{
\includegraphics[width=0.75\textwidth]{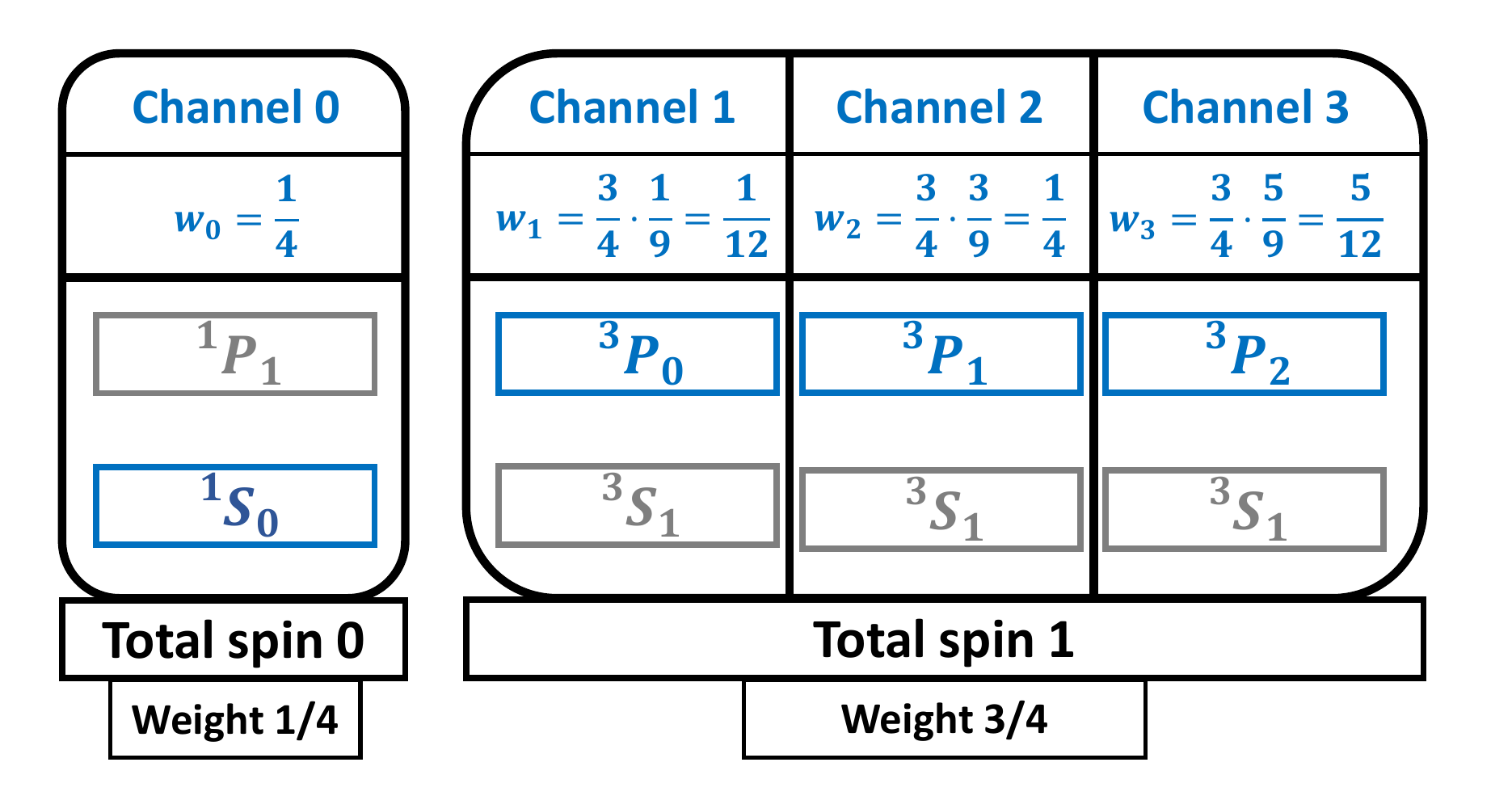}
}
\caption
{An example for the CATS input when both the s- and p-wave interactions are included. There are four distinct interaction channels: 
$\psi_0=\psi_{1S0}+\psi_{1P1}+...$, $\psi_1=\psi_{3S1}+\psi_{3P0}+...$, 
$\psi_2=\psi_{3S1}+\psi_{3P1}+...$ and $\psi_3=\psi_{3S1}+\psi_{3P2}+...$. 
The dots represent all partial waves beyond the $p$-wave, which are assumed to have no influence on $C(k)$ and are thus considered irrelevant. 
The grayed out states are not allowed in the case of p-p correlation, but can be included if needed, e.g. for p-$\mathrm{\Lambda}$ correlations. 
}
\label{fig:Channels}
\end{figure*}
\end{center}

\end{document}